\documentclass[preprint,11pt]{elsarticle}

 \usepackage{epsfig}

\usepackage{amssymb}
\usepackage{graphics,amssymb,amsmath}
\usepackage{graphicx}
\usepackage{subfigure}
\newproof{pf}{Proof}
\newdefinition{rmk}{Remark}
\newtheorem{thm}{Theorem}
\newdefinition{prop}{Proposition}
\biboptions{sort&compress}
\journal{Mathematics and Computers in Simulation}

\begin{document}

\begin{frontmatter}



\title{Exponentiated Weibull-Poisson distribution: model, properties and applications}


\author{Eisa Mahmoudi\corref{cor1}}
\ead{emahmoudi@yazduni.ac.ir}
\author{Afsaneh Sepahdar}
%\ead[url]{http://www.elsevier.com}

\cortext[cor1]{Corresponding author}

\address{Department of Statistics, Yazd University,
P.O. Box 89175-741, Yazd, Iran}
%\address[rvt]{Department of Statistics, Yazd University,
 %P.O. Box 89175-741, Yazd, Iran}

\begin{abstract}
In this paper we propose a new four-parameters distribution with
increasing, decreasing, bathtub-shaped and unimodal failure rate,
called as the exponentiated Weibull-Poisson (EWP) distribution. The
new distribution arises on a latent complementary risk problem base
and is obtained by compounding exponentiated Weibull (EW) and
Poisson distributions. This distribution contains several lifetime
sub-models such as: generalized exponential-Poisson (GEP),
complementary Weibull-Poisson (CWP), complementary
exponential-Poisson (CEP), exponentiated Rayleigh-Poisson (ERP) and
Rayleigh-Poisson (RP) distributions.

 We obtain several properties of the new distribution
such as its probability density function, its reliability and
failure rate functions, quantiles and moments. The maximum
likelihood estimation procedure via a EM-algorithm is presented in
this paper. Sub-models of the EWP distribution are studied in
details. In the end, Applications to two real data sets are given to
show the flexibility and potentiality of the new distribution.

\end{abstract}


\end{frontmatter}


\section{Introduction}

%CCCCCCCCCCCCCCCCCCCCCCCCCCCCCCCCCCCCCCCCCCCCCCCCCCCCCCCCCCCCCCCCCCCCCCCCCCCCCCCCCCCCCC
%CCCCCCCCCCCCCCCCCCCCCCCCCCCCCCCCCCCCCCCCCCCCCCCCCCCCCCCCCCCCCCCCCCCCCCCCCCCCCCCCCCCCCCCC
%The complementary risk problem arises in several areas, such as
%public health, actuarial science, biomedical studies, demography and
%industrial reliability.

The Weibull and EW distributions provide a simple and close form
solution to many problems in lifetime and reliability. However, they
do not provide a reasonable parametric fit for some practical
applications, for example the Weibull distribution is not useful for
modeling phenomenon with non-monotone failure rates. The
bathtub-shaped and the unimodal failure rates which are common in
reliability and biological studies cannot be modeled by the Weibull
distribution.

Recently, attempts have been made to define new families of
probability distributions that extend well-known families of
distributions and at the same time provide great flexibility in
modeling data in practice. The exponential-geometric (EG),
exponential-Poisson (EP), exponential-logarithmic (EL),
exponential-power series (EPS), Weibull-geometric (WG) and
Weibull-power series (WPS) distributions were introduced and studied
by Adamidis and Loukas \cite{Adamidis }, Kus \cite{Kus }, Tahmasbi
and Rezaei \cite{Tahmasbi }, Chahkandi and Ganjali \cite{Chahkandi},
Barreto-Souza et al. \cite{Barreto-Souza2011 } and Morais and
Barreto-Souza et al. \cite{Morais }, respectively.

Barreto-Souza and Cribari-Neto \cite{Barreto-Souza2009 } and Louzada
et al. \cite{Louzada} introduced the exponentiated
exponential-Poisson (EEP) and the complementary
exponential-geometric (CEG) distributions where the EEP is the
generalization of the EP distribution and the CEG is complementary
to the EG model proposed by Adamidis and Loukas \cite{Adamidis }.
Recently, Cancho et al. \cite{Cancho } introduced the two-parameter
Poisson-exponential (PE) lifetime distribution with increasing
failure rate. Mahmoudi and Jafari \cite{Mahmoudi2011a } introduced
the generalized exponential-power series (GEPS) distribution by
compounding the generalized exponential (GE) distribution with the
power series distribution. Also exponentiated Weibull-logarithmic
(EWL), exponentiated Weibull-geometric (EWG) and exponentiated
Weibull-power series (EWP) distributions has been introduced and
analyzed by Mahmoudi and Sepahdar \cite{Mahmoudi2011b} and Mahmoudi
and Shiran \cite{Mahmoudi2011c,Mahmoudi2011d}.

In this paper, we propose a new four-parameters distribution,
referred to as the EWP distribution, which contains as special
sub-models the GEP, CWP, CEP, ERP and RP distributions, among
others. The main reasons for introducing the EWP distribution are:
(i) This distribution due to its flexibility in accommodating
different forms of the risk function is an important model that can
be used in a variety of problems in modeling lifetime data. (ii)
This distribution is a suitable model in a complementary risk
problem base (Basu and Klein, \cite{Basu }) in presence of latent
risks, in the sense that there is no information about which factor
was responsible for the component failure and only the maximum
lifetime value among all risks is observed. (iii) It provides a
reasonable parametric fit to skewed data that cannot be properly
fitted by other distributions and is a suitable model in several
areas such as public health, actuarial science, biomedical studies,
demography and industrial reliability.

The paper is organized as follows. In Section 2, we review the EW
distribution and its properties. In Section 3, we define the EWP
distribution. The density, survival and hazard rate functions and
some of their properties are given in this section. We derive
quantiles and moments of the EWP distribution in Section 4.
R\'{e}nyi and Shannon entropies of the EWP distribution are given in
Section 5. Section 6 provides the moments of order statistics of the
EWP distribution. Residual and reverse residual life functions of
the EWP distribution are discussed in Section 7. In Section 8 we
explain probability weighted moments. Mean deviations from the mean
and median are obtained in Section 9. Section 10 is devoted to the
Bonferroni and Lorenz curves of the EWP distribution. Estimation of
the parameters by maximum likelihood via a EM-algorithm and
inference for large sample are presented in Section 11. In Section
12, we studied some special sub-models of the EWP distribution.
Applications to real data sets are given in Section 13 and
conclusions are provided in Section 14.

%CCCCCCCCCCCCCCCCCCCCCCCCCCCCCCCCCCCCCCCCCCCCCCCCCCCCCCCCCCCCCCCCCCCCCCCCCCCCCCCCC
%CCCCCCCCCCCCCCCCCCCCCCCCCCCCCCCCCCCCCCCCCCCCCCCCCCCCCCCCCCCCCCCCCCCCCCCCCCCCCCCCCC
%CCCCCCCCCCCCCCCCCCCCCCCCCCCCCCCCCCCCCCCCCCCCCCCCCCCCCCCCCCCCCCCCCCCCCCCCCCCCCCCCCCCCCCC

\section{ EW distribution: A brief review }

The EW distribution, introduced by Mudholkar and Srivastava
\cite{Mudholkar1993 } as extension of the Weibull family, contains
distributions with bathtub-shaped and unimodal failure rates besides
a broader class of monotone failure rates. The applications of the
EW distribution in reliability and survival studies were illustrated
by Mudholkar et al. \cite{Mudholkar1995 }, Mudholkar and Huston
\cite{Mudholkar1996 }, Gupta and Kundu \cite{Gupta2001 }, Nassar and
Eissa \cite{Nassar} and Choudhury \cite{Choudhury}.

The random variable $X$ has an EW distribution if its cumulative
distribution function (cdf) takes the form
\begin{equation}\label{cdf ExW}
F_{X}(x)=\left(1-e^{-(\beta x)^{\gamma}}\right)^{\alpha},~~x>0,
\end{equation}
where $\alpha>0$, $\beta>0$ and $\gamma>0$. The corresponding
probability density function (pdf) is
\begin{equation}\label{pdf ExW}
f_X(x)=\alpha\gamma\beta^{\gamma}x^{\gamma-1}e^{-(\beta
x)^{\gamma}}\left(1-e^{-(\beta x)^{\gamma}}\right)^{\alpha-1}.
\end{equation}
The survival and hazard rate functions of the EW distribution are
\begin{equation*}
S(x)=1-\left(1-e^{-(\beta x)^{\gamma}}\right)^{\alpha},
\end{equation*}
and
\begin{equation*}
h(x)= \alpha\gamma\beta^{\gamma}x^{\gamma-1}e^{-(\beta
x)^{\gamma}}\left(1-e^{-(\beta
x)^{\gamma}}\right)^{\alpha-1}\Big\{\left[1-\left(1-e^{-(\beta
x)^{\gamma}}\right)^{\alpha}\right]\Big\}^{-1},
\end{equation*}
respectively. The $k$th moment about zero of the EW distribution is
given by
\begin{equation*}\label{mean EW}
E(X^{k})=\alpha\beta^{-k}\Gamma\left(\frac{k}{\gamma}+1\right)\sum_{j=0}^{\infty}(-1)^j
{\alpha-1 \choose j}(j+1)^{-(\frac{k}{\gamma}+1)}.
\end{equation*}
Note that for positive integer values of $\alpha$, the index $j$ in
previous sum stops at $\alpha-1$ and the above expression takes the
closed form
\begin{equation*}
E(X^{k})=\alpha\beta^{-k}\Gamma\left(\frac{k}{\gamma}+1\right)A_{k}(\gamma),
\end{equation*}
when
\begin{equation*}\label{v.p}
A_{k}(\gamma)=1+\sum_{j=1}^{\alpha-1}(-1)^j {\alpha-1 \choose
j}(j+1)^{-(\frac{k}{\gamma}+1)},~~~k=1,2,3,\cdots,
\end{equation*}
where $\Gamma(\frac{k}{\gamma}+1)$ denotes the gamma function (see,
Nassar and Eissa \cite{Nassar} for more details).

\section{ The EWP distribution }
Suppose that the random variable $X$ has the EW distribution where
its cdf and pdf are given in (1) and (2). Given $N$, let
$X_{1},\cdots,X_{N}$ be independent and identify distributed random
variables from EW distribution. Let $N$ is distributed according to
zero truncated Poisson distribution with pdf
\begin{equation*}
P(N=n)=\frac{e^{-\theta}\theta^{n}}{n!(1-e^{-\theta})},~n=1,2
,\cdots,~\theta>0.
\end{equation*}
Let $Y=\max(X_{1},\cdots,X_{N})$, then the cdf of $Y|N=n$ is given
by
\begin{equation*}\label{dist y given N}
F_{Y|N=n}(y)=[1 - e^{-(‎\beta y‎)^{‎\gamma‎}}]^{n
‎\alpha‎},
\end{equation*}
which is EW distribution with parameters $n\alpha$, $\beta$ and
$\gamma$. The EWP distribution, denote by
EWP$(\alpha,\beta,\gamma,\theta)$, is defined by the marginal cdf of
$Y$, i.e.
\begin{equation}\label{cdf EWP}
F(y)= ‎\frac{e^{‎\theta (1 -e^{-(‎\beta y‎)^{‎\gamma‎}}
)^{‎\alpha‎}}‎ -1}{e^{‎\theta‎}- 1}‎.
\end{equation}
This new distribution includes some sub-models such as CEP, GEP,
GWP, ERP and RP distributions.
\\The pdf of EWP distribution is given by
\begin{equation}\label{pdf EWP}
f(y)=‎\frac{‎\alpha‎ \gamma‎ \theta‎‎‎ ‎\beta ^{‎\gamma‎} y^{‎\gamma -1‎}‎}{e^{‎\theta‎} - 1}
 e^{-(‎\beta  y‎)^{‎\gamma‎}}(1 -e^{-(‎\beta y‎)^{‎\gamma‎}})^
 {‎\alpha -1‎} e^{‎\theta (1 -e^{-(‎\beta y‎)^{‎\gamma‎}} )^{‎\alpha‎}‎} ‎,
\end{equation}
where $\alpha,\beta,\gamma,\theta>0$.\\
The survival and hazard rate functions of EWP distribution are
given, respectively, by
\begin{equation}\label{survive}
S(y)=1 -‎\frac{e^{‎\theta (1 -e^{-(‎\beta
y‎)^{‎\gamma‎}} )^{‎\alpha‎}‎} -1}{e^{‎\theta‎}- 1}
,
\end{equation}
and
\begin{equation}\label{hazard}
h(y)=‎‎\frac{‎\alpha‎ \gamma‎ \theta‎‎‎ ‎\beta
^{‎\gamma‎} y^{‎\gamma -1‎} e^{-(‎\beta
y‎)^{‎\gamma‎}}(1 -e^{-(‎\beta
y)^{‎\gamma‎}})^{‎\alpha -1‎} e^{‎\theta (1 -e^{-(‎\beta
y)^{‎\gamma‎}} )^{‎\alpha‎}‎}}{e^{‎\theta‎} -
e^{‎\theta (1 -e^{-(‎\beta y‎)^{‎\gamma‎}}
)^{‎\alpha}}}‎‎.
\end{equation}
The plots of density and hazard rate functions of EWP distribution
for $\beta=1$ and different values $(\alpha,\gamma,\theta)$ are
given in Figs. 1 and 2, respectively.
\begin{figure}[t]
\centering
\includegraphics[scale=0.32]{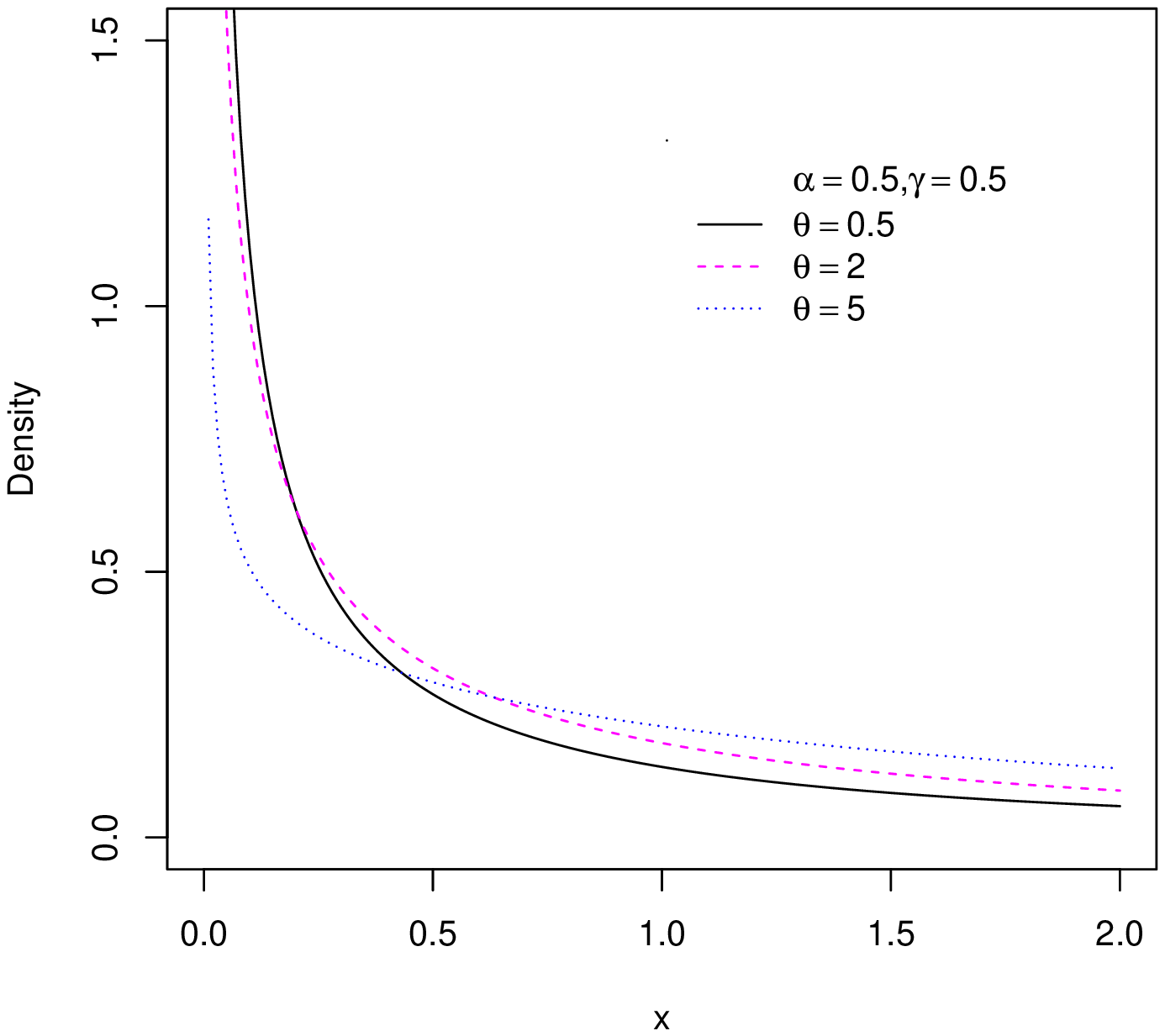}
\includegraphics[scale=0.32]{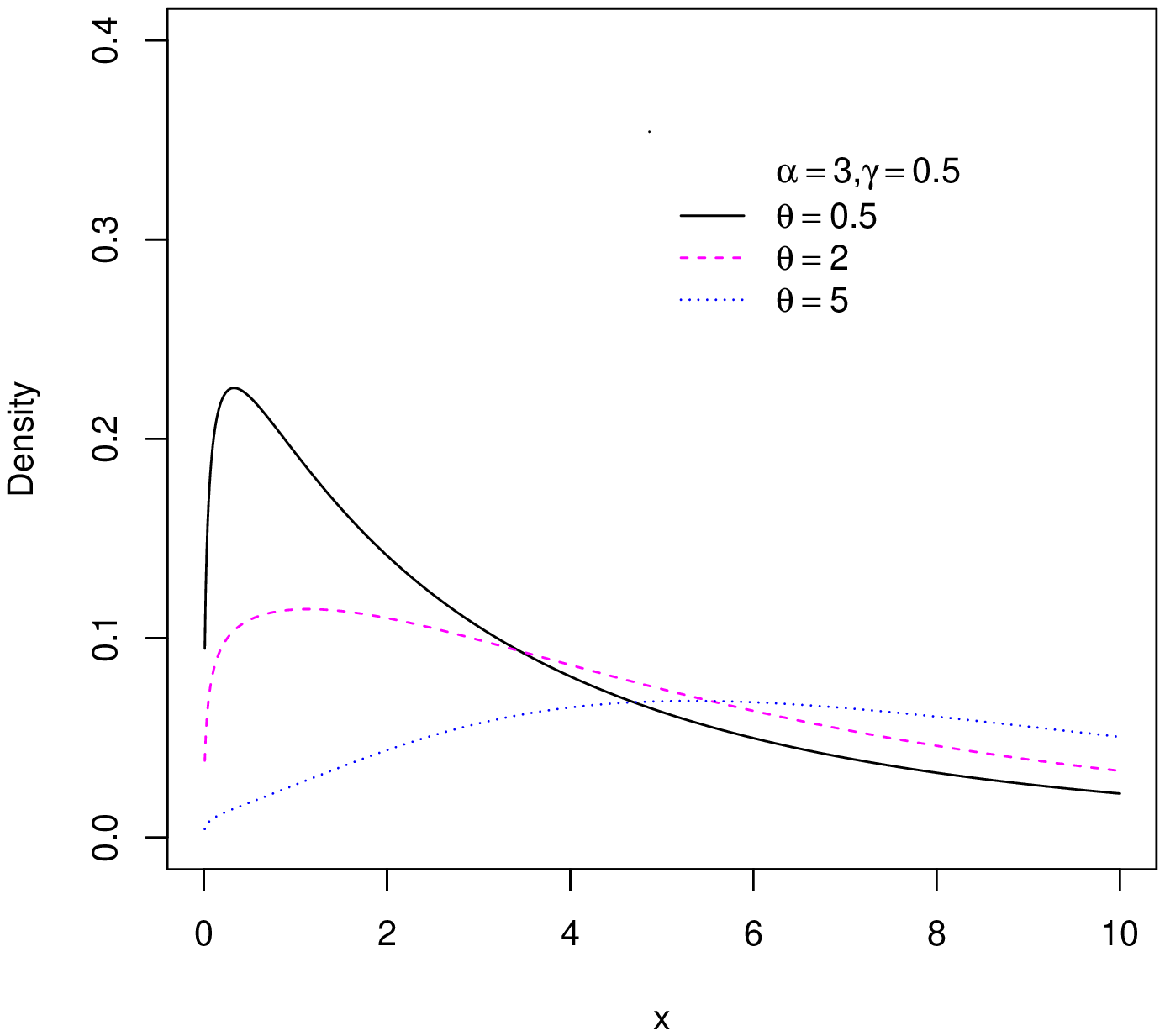}
\includegraphics[scale=0.32]{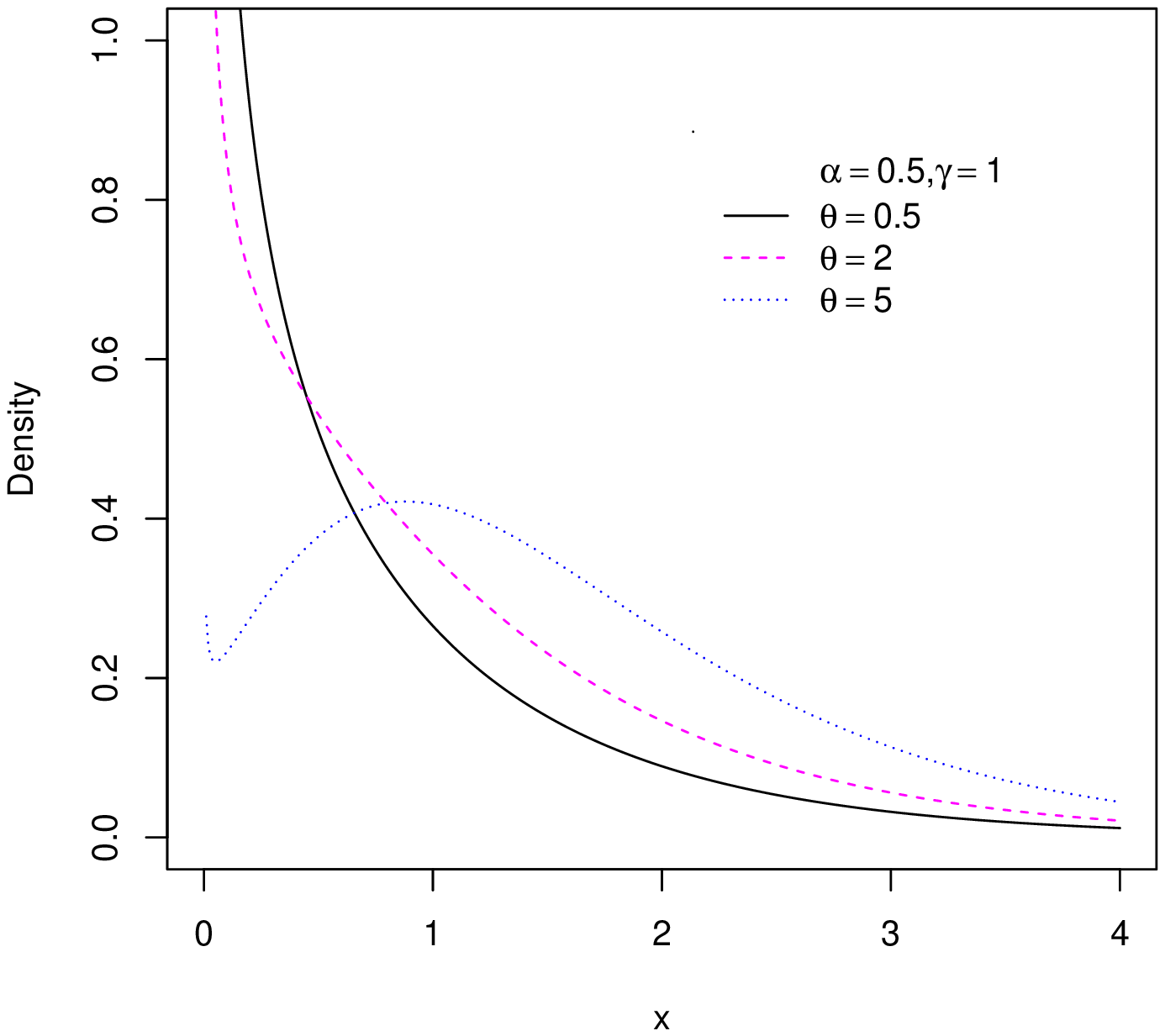}
\includegraphics[scale=0.32]{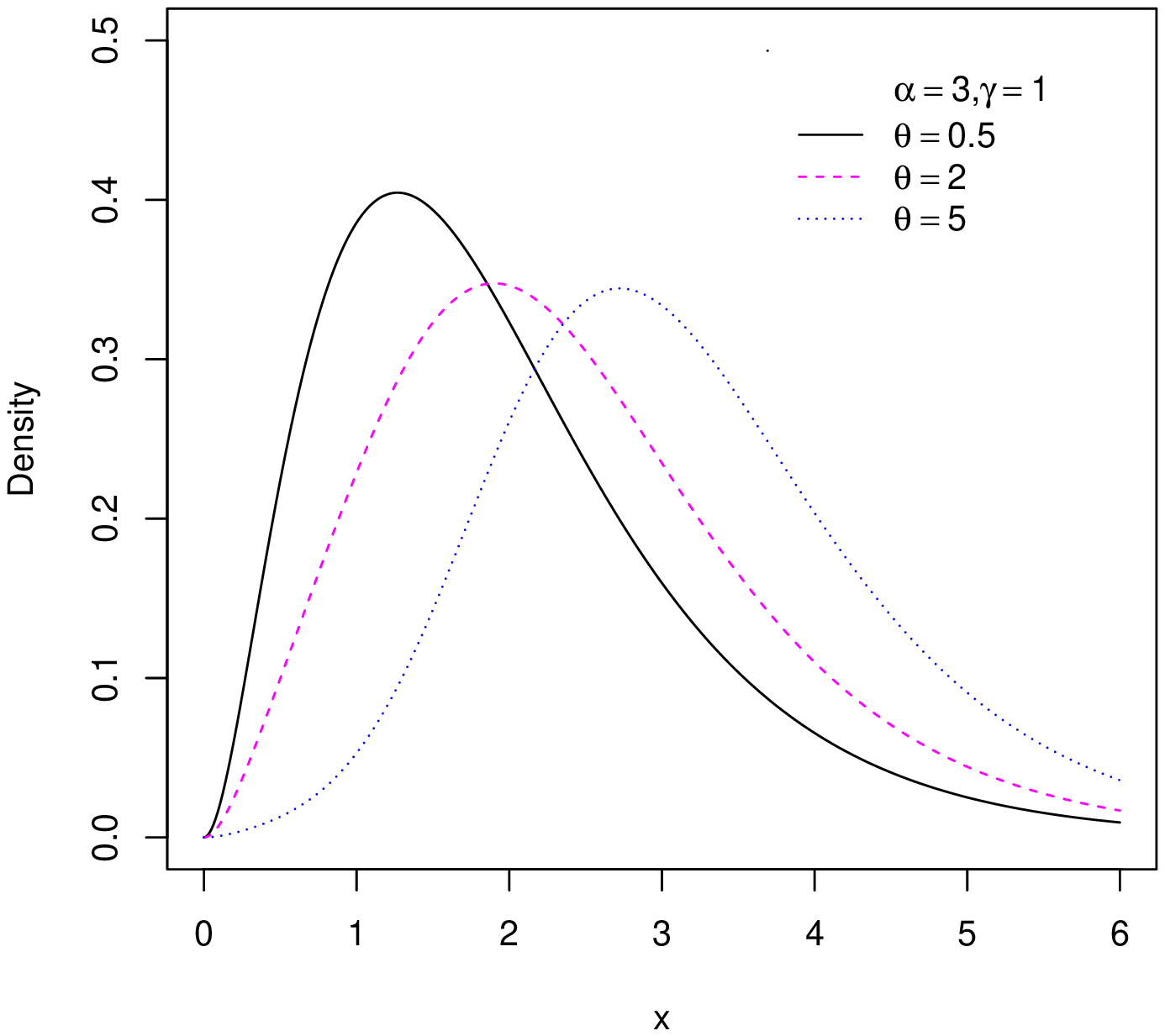}
\includegraphics[scale=0.32]{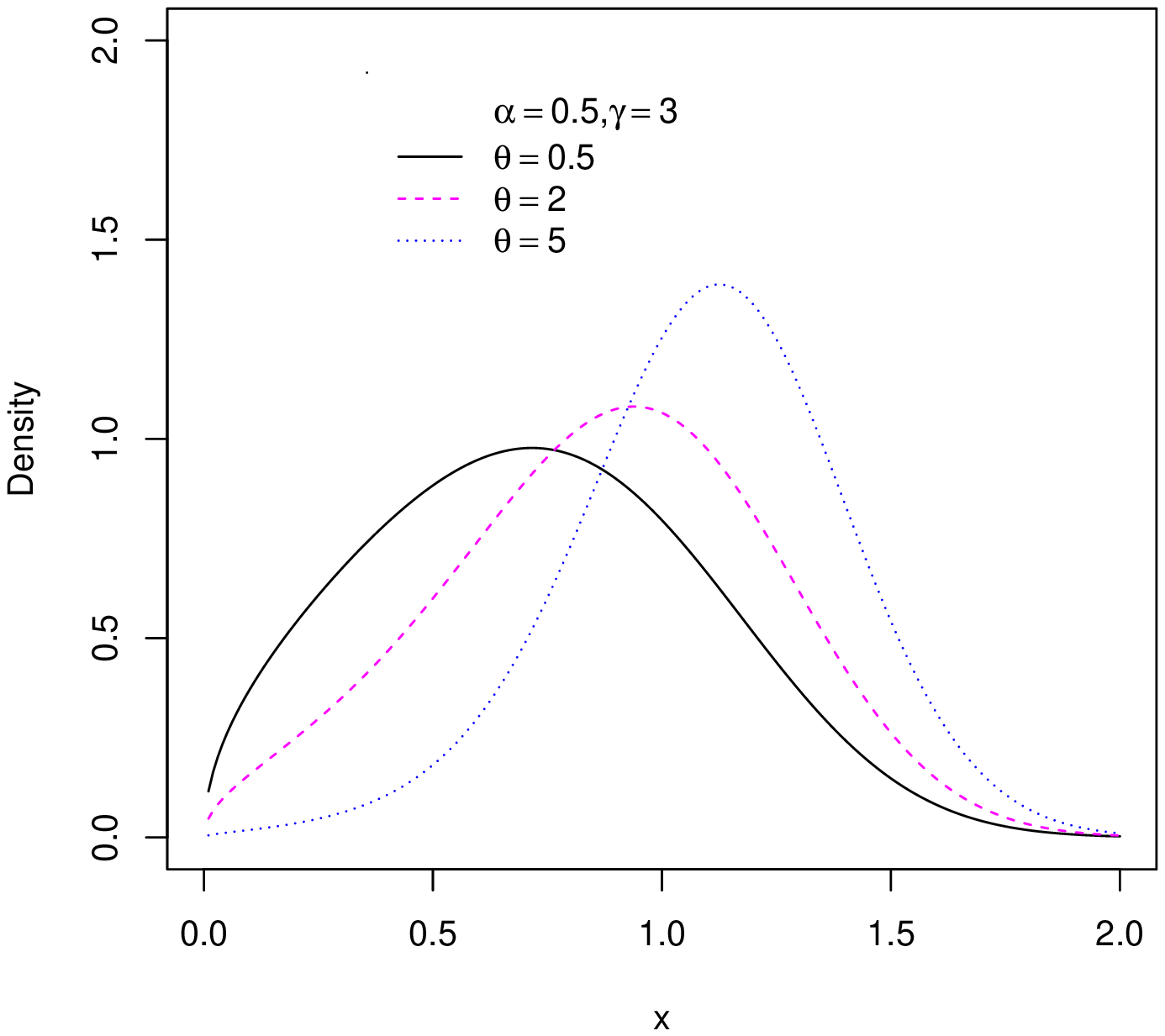}
\includegraphics[scale=0.32]{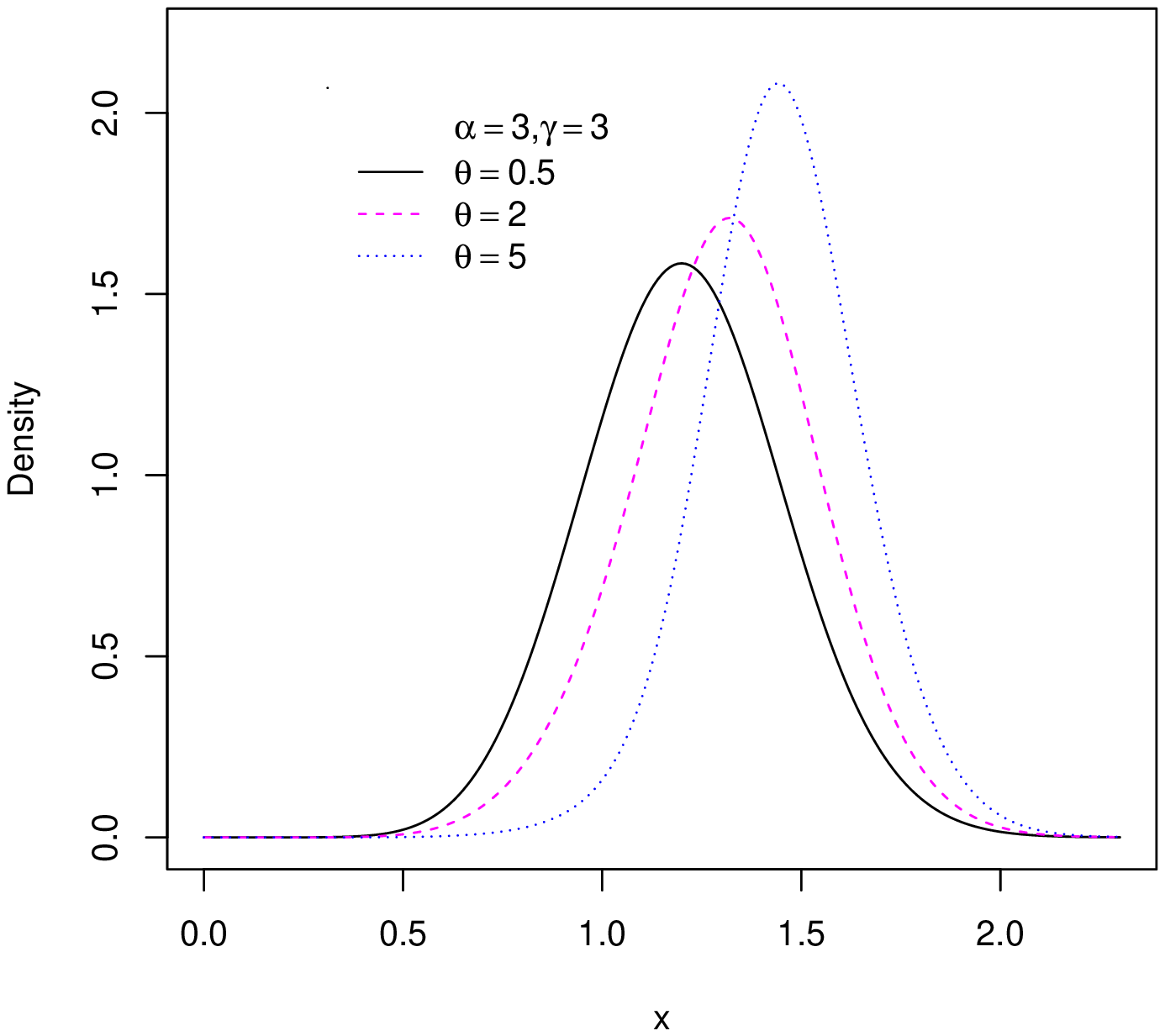}
\caption[]{Plots of density function of EWP distribution for
$\beta=1$ and different values $\alpha $, $\theta $ and $\gamma$.}
\end{figure}

\begin{figure}[t]
\centering
\includegraphics[scale=0.32]{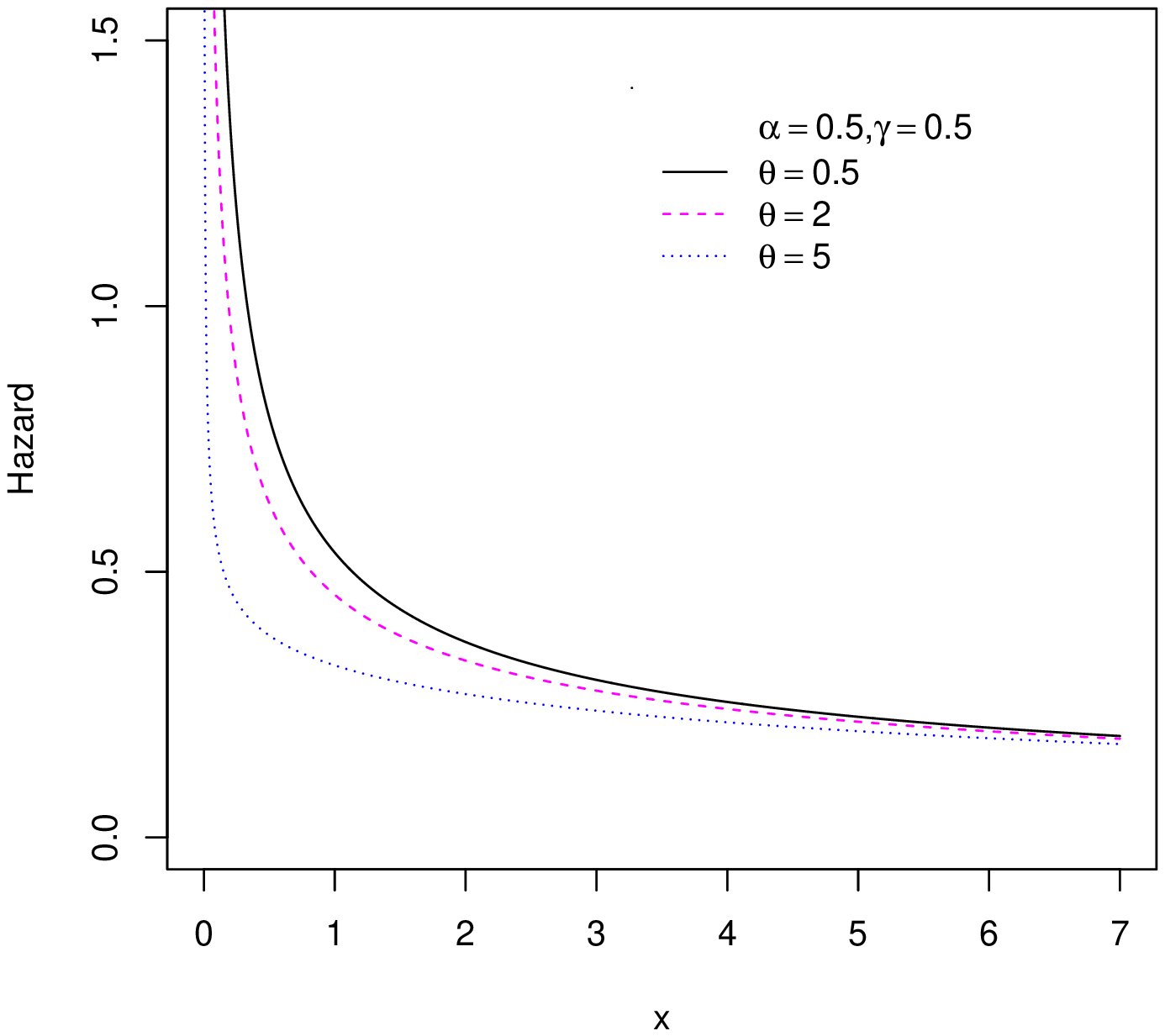}
\includegraphics[scale=0.32]{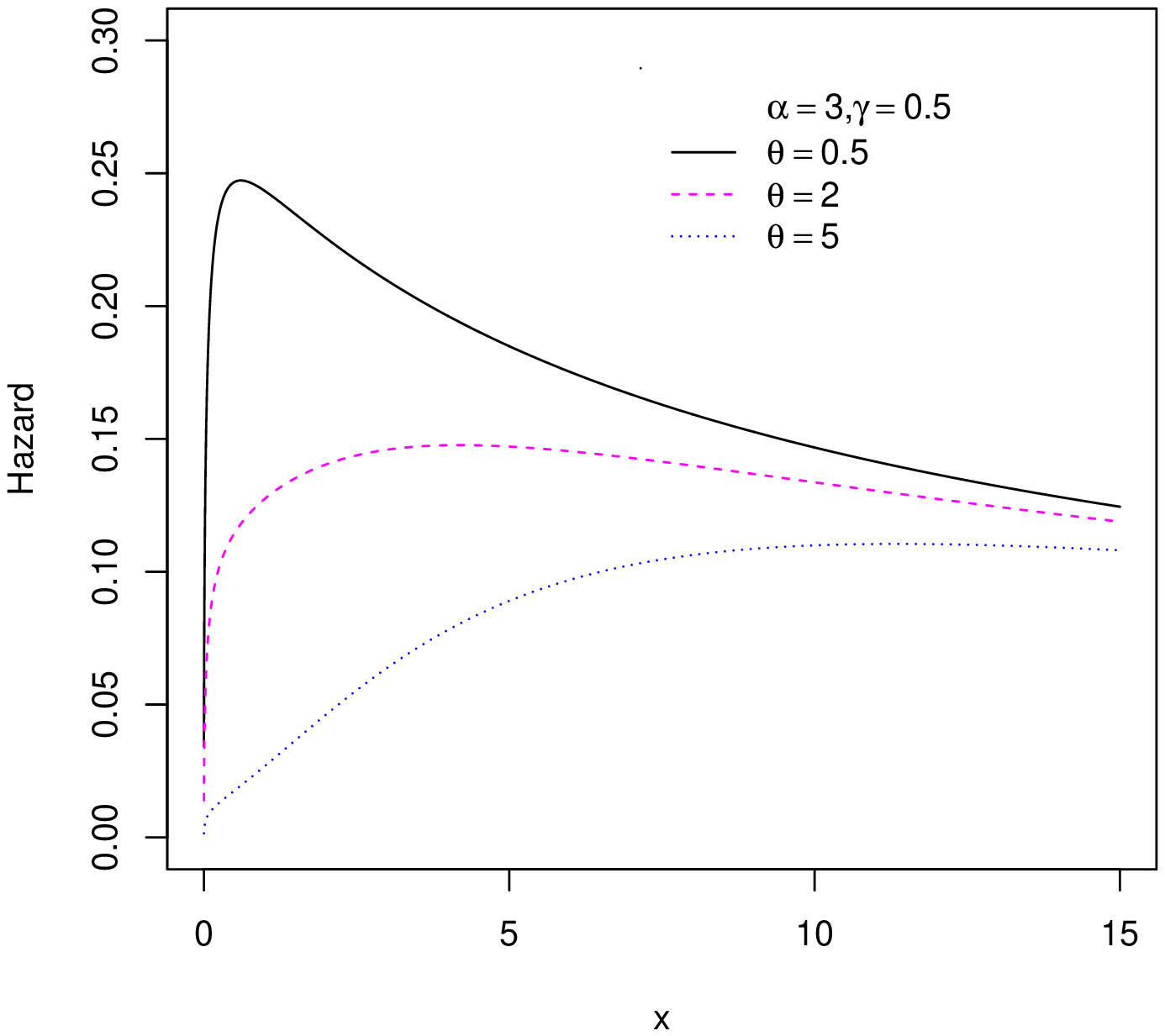}
\includegraphics[scale=0.32]{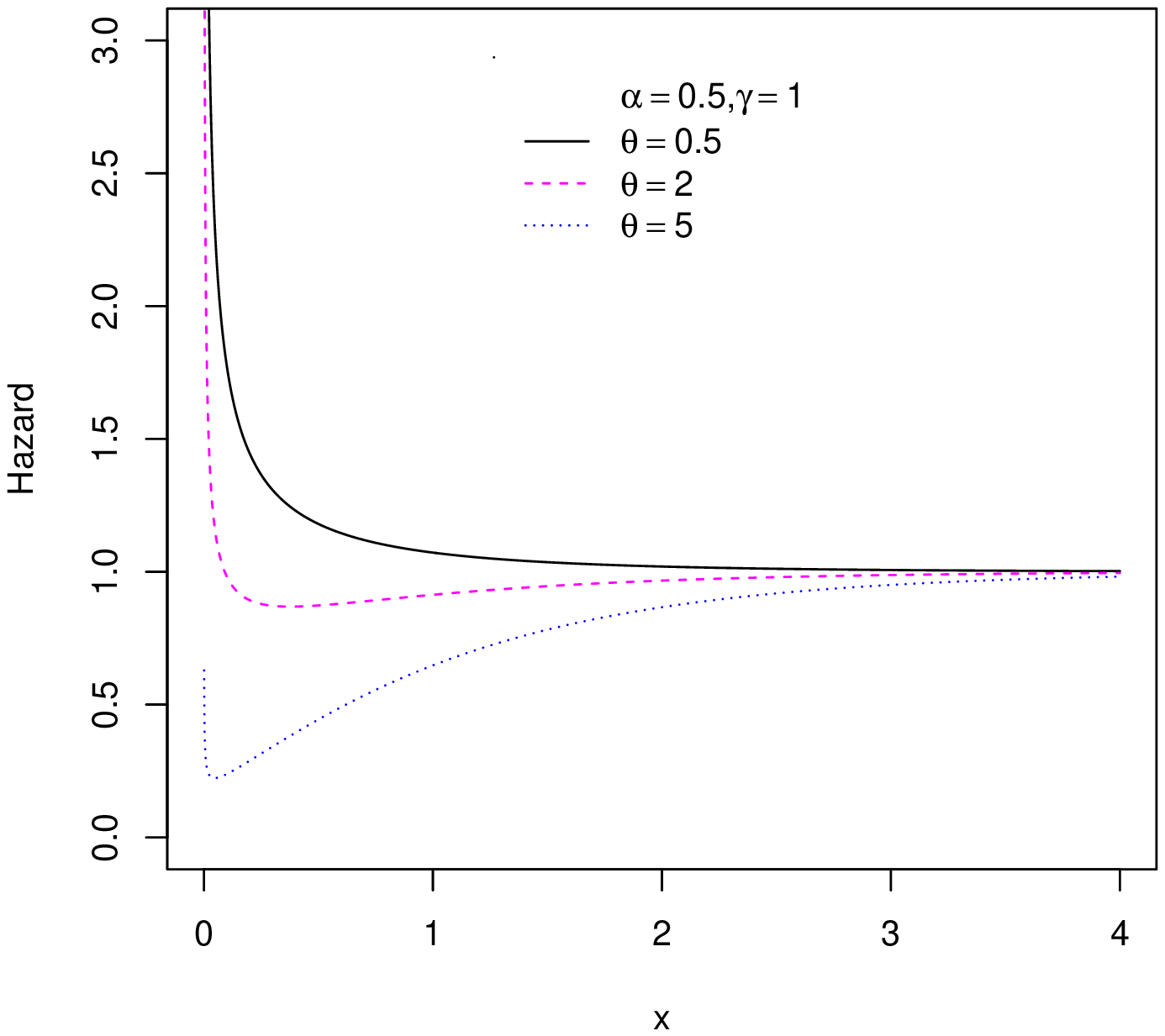}
\includegraphics[scale=0.32]{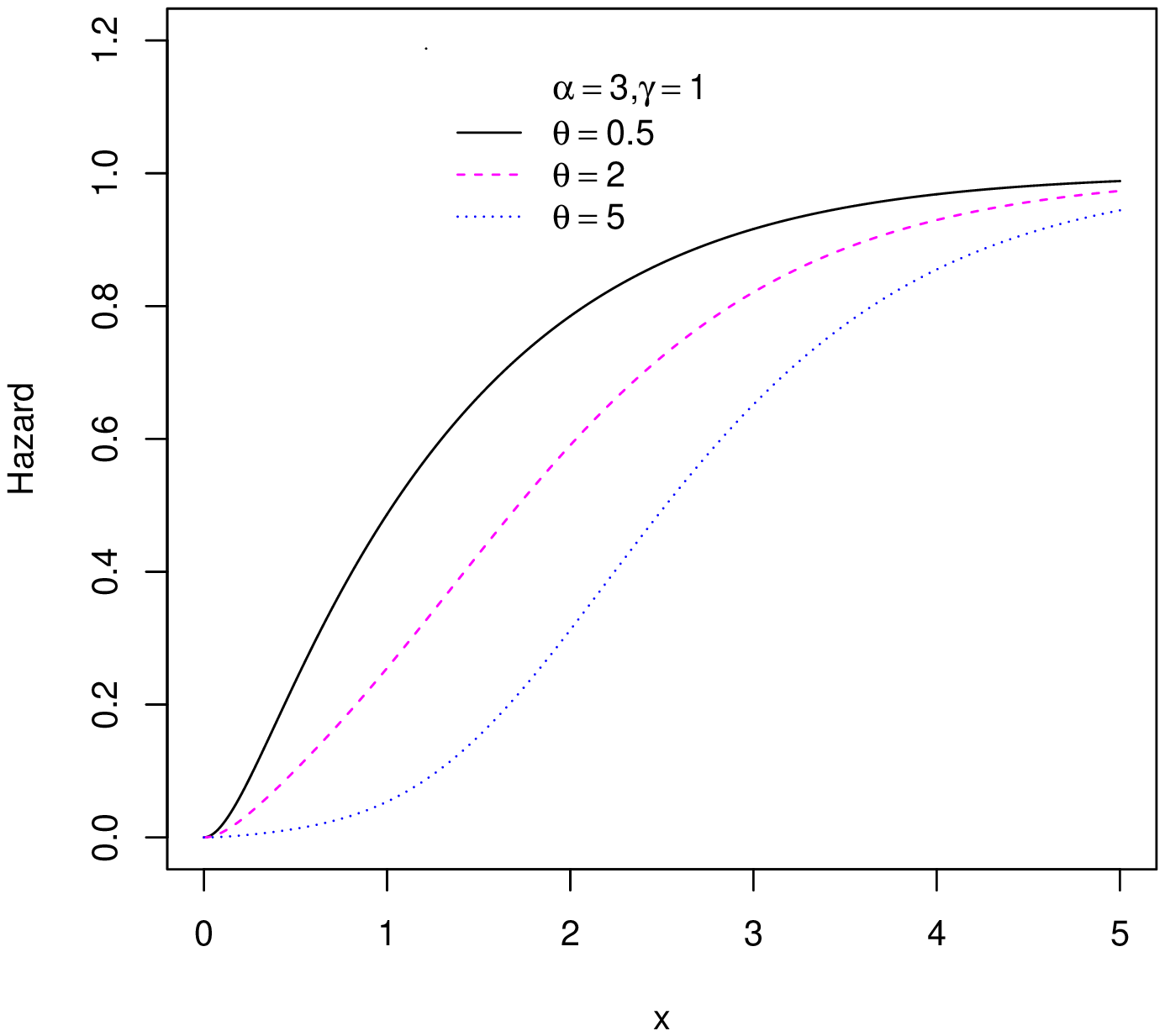}
\includegraphics[scale=0.32]{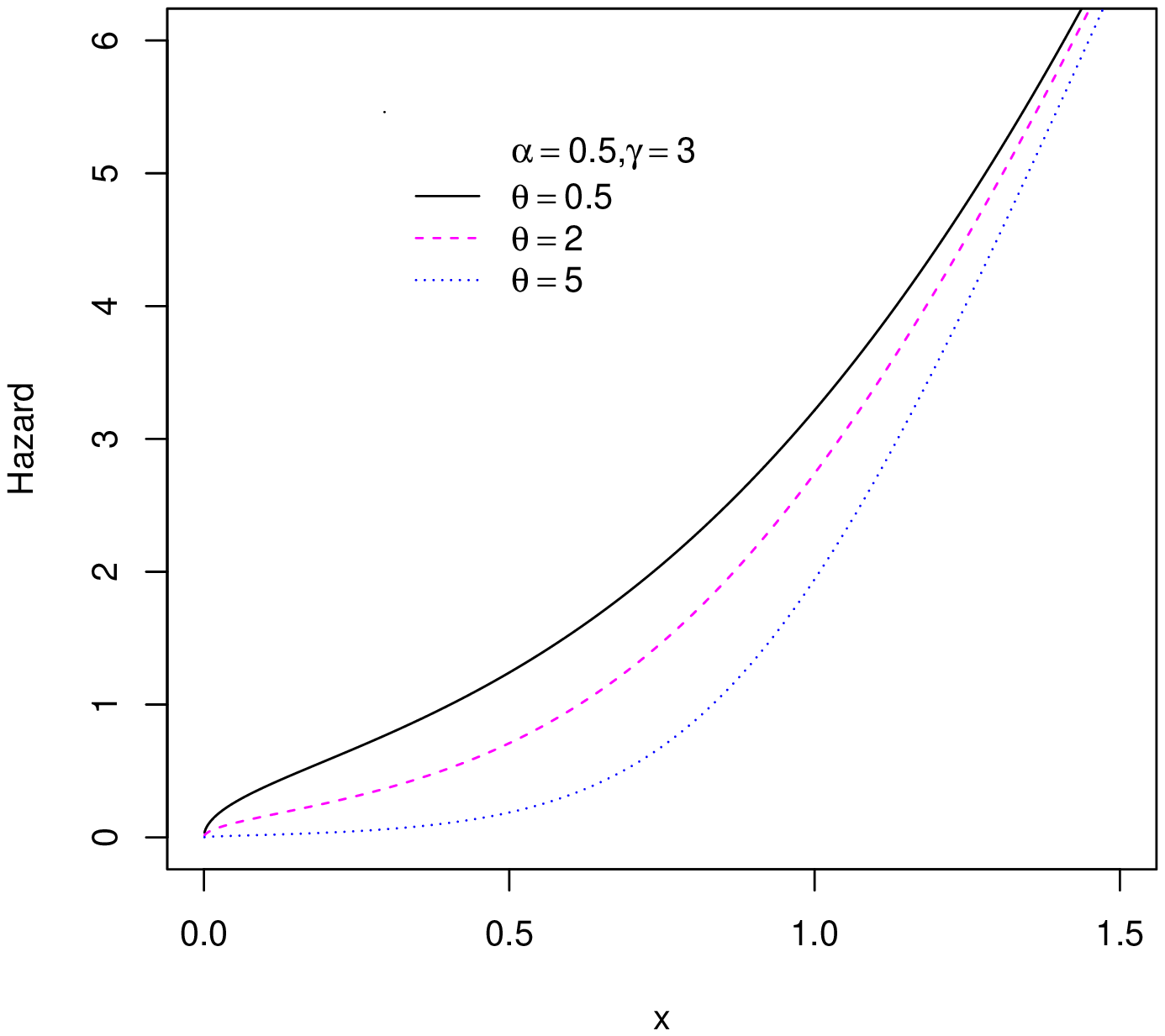}
\includegraphics[scale=0.32]{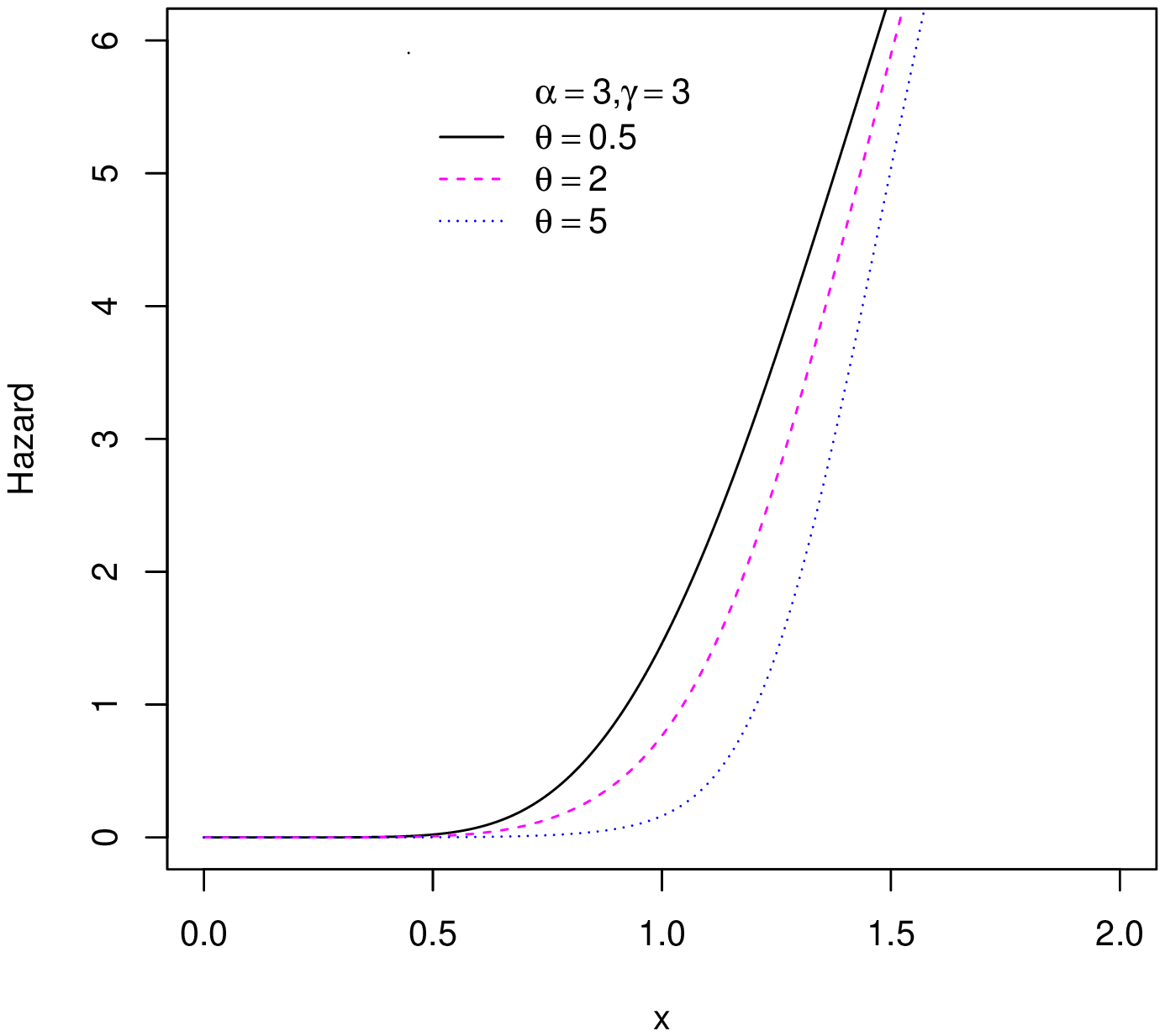}
\caption[]{Plots of hazard function of EWP distribution for
$\beta=1$ and different values $\alpha $, $\theta $ and $\gamma$.}
\end{figure}

\begin{prop}
The limiting distribution of EWP $(\alpha,\beta,\gamma,\theta)$
where $\theta\rightarrow 0^{+}$ is
\begin{equation*}
\lim_{\theta\rightarrow 0^{+}}F(y)=‎ (1 - e^ {- (‎\beta
y‎)^{‎\gamma‎}})^{‎\alpha‎}‎,
\end{equation*}
which is the cdf of EW distribution.
\end{prop}
\begin{prop}
The limiting behavior of hazard function of EWP distribution in
(\ref{hazard}) is\\
(i) for $0<\gamma<1$, $\lim_{y\rightarrow 0}h(y)=\left\{
\begin{array}{lc}
\infty  & 0<\alpha \leq1 \\
0 & \alpha >1, \end{array} \right.$ and $\lim_{y\rightarrow
\infty}h(y)=0.$\\
(ii) for $\gamma=1$, $\lim_{y\rightarrow 0}h(y)=\left\{
\begin{array}{lc}
\infty  & 0<\alpha <1 \\
\frac{\theta\beta}{e^{\theta}-1} & \alpha =1 \\
0 & \alpha>1, \end{array} \right.$ and $\lim_{y\rightarrow
\infty}h(y)=\beta.$\\
(iii) for $\gamma>1$, $\lim_{y\rightarrow 0}h(y)=0,$ for each value
$\alpha>0$ and $\lim_{y\rightarrow \infty}h(y)=\infty$.
\end{prop}
\begin{pf}
The proof is a forward calculation and is omitted.
\end{pf}

%The density of EWP distribution can be expressed as infinite linear
%combination of density of the biggest order statistic i.e.,
%\begin{equation*}
%f(y)=\sum^{\infty}_{n=1}(G(y))^{n}P(N=n)=\sum^{\infty}_{n=1}g_{X_{(n)}}(y)P(N=n),
%\end{equation*}
%in which $g_{X_{(n)}}(y)$ is the pdf of
%$X_{(n)}=\max(X_{1},\cdots,X_{n})$, where $X_{i}\sim
%EW(\alpha,\beta,\gamma)$ for $i=1,2,\cdots,n$.
%\end{prop}

\section{ Quantiles and moments of the EWP distribution }
Some of the most important features and characteristics of a
distribution can be studied through its moments and quantiles such
as tending, dispersion, skewness and kurtosis. Also, the quantiles
of a distribution can be used in data generation from a
distribution.

The $p$th quantile of the EWP distribution is given by
\begin{equation}\label{p quantile}
x_{p}=\frac{1}{‎\beta‎}‎ \left[-\log \left( 1-\left(
‎\frac{1}{‎\theta‎} \log \left( p\left(e^{‎\theta‎}
-1\right)
+1\right)‎\right)^{1/‎\alpha‎}\right)\right]^{1/‎\gamma‎},
\end{equation}
which is used for data generation from the EWP distribution.\\
Now we obtain the moment generating function of the EWP
distribution. Suppose that $Y\sim EWP(\alpha,\beta,\gamma,\theta)$
and $X_{(n)}=\max(X_{1},\cdots,X_{n})$, where $X_{i}\sim
EW(\alpha,\beta,\gamma)$ for $i=1,2,\cdots,n$,  then
\begin{equation}
\begin{array}[b]{ll}\label{mgfEWP}
M_{Y}(t)&=\sum^{\infty}_{n=1}P(N=n)M_{X_{(n)}}(t)\medskip \\
&=\sum^{\infty}_{n=1}P(N=n)\sum^{\infty}_{k=0}\frac{t^{k}}{k!}n\alpha\beta^{-k}\Gamma(1+\frac{k}{\gamma})\sum^{\infty}_{j=0}(-1)^{j}
{n\alpha-1 \choose j}(j+1)^{-(1+\frac{k}{\gamma})}\medskip \\
&=\frac{\alpha\theta}{(e^{\theta}-1)}\sum^{\infty}_{n=1}\sum^{\infty}_{k=0}\sum^{\infty}_{j=0}(-1)^{j}\frac{\theta^{n-1}}{(n-1)!}\frac{t^{k}\beta^{-k}}{k!}
{n\alpha-1 \choose
j}(j+1)^{-(1+\frac{k}{\gamma})}\Gamma(1+\frac{k}{\gamma}).
\end{array}
\end{equation}
One can use $M_{Y}(t)$ to obtain the $k$th moment about zero of the
EWP distribution. We have
\begin{equation}\label{meankEWP}
\begin{array}[b]{ll}
E(Y^{k})&=\sum^{\infty}_{n=1}P(N=n)E(X^{k}_{(n)})  \medskip\\
&  =
\frac{\alpha\theta}{\beta^{k}(e^{\theta}-1)}\sum^{\infty}_{n=1}\frac{\theta^{n-1}}{(n-1)!}\alpha\beta^{-k}\sum^{\infty}_{j=0}(-1)^{j}
{n\alpha-1 \choose j}(j+1)^{-(1+\frac{k}{\gamma})}\Gamma(1+\frac{k}{\gamma})   \medskip\\
&  =
\frac{\alpha\theta\Gamma(1+\frac{k}{\gamma})}{\beta^{k}(e^{\theta}-1)}\sum^{\infty}_{n=1}\sum^{\infty}_{j=0}(-1)^{j}{n\alpha-1
\choose j}\frac{\theta^{n-1}}{(n-1)!}(j+1)^{-(1+\frac{k}{\gamma})}.
\end{array}
\end{equation}
Using the change of variable $i=n-1,$ another equivalent formula for
$E(Y^{k})$ is given by
\begin{equation}\label{meankEWP2}
E(Y^{k})=\frac{\alpha\theta\Gamma(1+\frac{k}{\gamma})}{\beta^{k}(e^{\theta}-1)}\sum^{\infty}_{i=0}\sum^{\infty}_{j=0}(-1)^{j}{\alpha(i+1)-1
\choose j}\frac{\theta^{i}}{i!}(j+1)^{-(1+\frac{k}{\gamma})}.
\end{equation}
 The mean and variance of the EWP distribution
are given, respectively, by
\begin{equation}\label{meanEWP}
E(Y)=\frac{\alpha \theta\Gamma (1+\frac{1}{\gamma}
)}{\beta(e^{\theta}-1)}\sum^{\infty}_{n=1}\sum^{\infty}_{j=0}(-
1)^{j} \frac{\theta^{n-1}}{(n-1)!}{n\alpha - 1 \choose j}
(j+1)^{-(1+\frac{1}{\gamma})} ,
\end{equation}
and
\begin{equation}\label{varEWP}
\begin{array}[b]{ll}
Var(Y)=\frac{\alpha \theta\Gamma (1+\frac{2}{\gamma} )}{\beta^{
2}(e^{\theta}-1)}\sum^{\infty}_{n=1}\sum^{\infty}_{j=0}(- 1)^{j}
\frac{\theta^{n-1}}{(n-1)!}{n\alpha - 1 \choose j}
(j+1)^{-(1+\frac{2}{\gamma})}-E^2(Y),
\end{array}
\end{equation}
where $E(Y)$ is given in Eq. (\ref{meanEWP}). Note that for positive
integer values of $\alpha$, the index $j$ in Eqs.
(\ref{mgfEWP})-(\ref{varEWP}) stops at $n\alpha-1$.

\section{ R\'{e}nyi and Shannon entropies}
If $X$ is a random variable having an absolutely continuous
cumulative distribution function $F(x)$ and probability distribution
function $f(x)$, then the basic uncertainty measure for distribution
$F$ (called the entropy of $F$) is defined as $H(x)=E[-\log(f(X))]$.
Statistical entropy is a probabilistic measure of uncertainty or
ignorance about the outcome of a random experiment, and is a measure
of a reduction in that uncertainty. Numerous entropy and information
indices, among them the R\'{e}nyi entropy, have been developed and
used in various disciplines and contexts. Information theoretic
principles and methods have become integral parts of probability and
statistics and have been applied in various branches of statistics
and related fields.

Entropy has been used in various situations in science and
engineering. The entropy of a random variable $Y$ is a measure of
variation of the uncertainty. For a random variable with the pdf
$f$, the R\'{e}nyi entropy is defined by
$I_{R}(r)=\frac{1}{1-r}\log\{\int_{\mathbb{R}}f^{r}(y)dy\}$, for
$r>0$ and $r\neq 1$. Using the power series expansion
$(1-z)^{\alpha}=\sum_{j=0}^{\infty}(-1)^{j}{\alpha \choose j}z^j$
and change of variable $(\beta y)^\gamma=u$ gives
\begin{equation*}
\begin{array}[b]{ll}
\int^{\infty}_{0} f^{r}(y)dy=\big ( ‎\frac{‎\alpha ‎\gamma ‎\theta ‎\beta ^{‎\gamma‎}‎‎‎‎}{e^{ ‎\theta -1‎}}‎\big)^{r}\sum^{\infty}_{j=0}‎\frac{(r‎\theta‎)^j}{j!}‎ \sum^{\infty}_{k=0} (-1)^{k}{‎\alpha(r+j)‎ -r\choose{k}}\int_{0}^{\infty}y^{ r(‎\gamma-1‎)} e^{-(r+k)(‎\beta y‎)^{‎\gamma‎}}dy.
 \end{array}
\end{equation*}
Using formula $\int_{0}^{\infty}y^{ r(‎\gamma-1‎)}
e^{-(r+k)(‎\beta
y‎)^{‎\gamma‎}}dy=\Gamma(‎\frac{r(‎\gamma-1‎)+1}{‎\gamma‎}‎)‎
\big( ‎\gamma ‎\beta ^{r(‎\gamma-1‎)+1)‎}\big)^{-1} \big(
r+k\big)^{-(‎‎\frac{r(‎\gamma-1‎)+1}{‎\gamma‎}‎)}$, we
have
\begin{equation*}
\begin{array}[b]{ll}
‎ \int^{\infty}_{0} f^{r}(y)dy& = ‎\frac{‎(\alpha ‎\theta)^{r}(‎\gamma  ‎\beta )^{‎r-1‎}‎‎‎‎}
{(e^{ ‎\theta -1‎})^{r}}‎\Gamma(‎\frac{r(‎\gamma-1‎)+1}{‎\gamma‎}‎)‎
\sum^{\infty}_{j=0}‎\sum^{\infty}_{k=0}\frac{ (-1)^{k}(r‎\theta‎)^j}{j!}‎
{‎\alpha(r+j)‎ -r\choose{k}} \big( r+k\big)^{-(‎‎\frac{r(‎\gamma-1‎)+1}{‎\gamma‎}‎)}.
 \end{array}
\end{equation*}
Thus, according to the definition of R\'{e}nyi entropy we have
\begin{equation*}
I_{R}(r) =\frac{1}{1-r} \log‎\left [‎‎\frac{‎(\alpha ‎\theta)^{r}(‎\gamma  ‎\beta )^{‎r-1‎}‎‎‎‎}{(e^{ ‎\theta -1‎})^{r}}‎\Gamma(‎\frac{r(‎\gamma-1‎)+1}{‎\gamma‎}‎)‎  \sum^{\infty}_{j=0}‎\sum^{\infty}_{k=0}\frac{ (-1)^{k}(r‎\theta‎)^j}{j!}‎ {‎\alpha(r+j)‎ -r\choose{k}} \big( r+k\big)^{-(‎‎\frac{r(‎\gamma-1‎)+1}{‎\gamma‎}‎)}\right].
\end{equation*}
The Shannon entropy is defined by $E[-\log [f(Y)]]$. This is a
special case derived from  $\lim_{r\rightarrow 1}I_{R}(r)$.

\section{ Moments of order statistics}

Order statistics make their appearance in many areas of statistical
theory and practice. Moments of order statistics play an important
role in quality control testing and reliability, where a
practitioner needs to predict the failure of future items based on
the times of a few early failures. These predictors are often based
on moments of order statistics. We now derive an explicit expression
for the density function and cumulative distribution function of the
$r$th order statistic $Y_{r:n}$, in a random sample of size $n$ from
the EWP distribution.

Let the random variable $Y_{r:n}$ be the $r$th order statistic
$(Y_{1:n}\leq Y_{2:n}\leq \cdots\leq Y_{n:n})$ in a sample of size
$n$ from a distribution with pdf $f(.)$ and cdf $F(.)$,
respectively. The pdf of $Y_{r:n}$ for $r=1,\cdots,n$ is given by
\begin{equation}\label{ordr EWP}
f_{r:n}(y)=\frac{1}{B(r,n-r+1)}f(y)[F(y)]^{r-1}[1-F(y)]^{n-r},~~~y>0.
\end{equation}
Substituting from (\ref{cdf EWP}) and (\ref{pdf EWP}) into
(\ref{ordr EWP}) and using the binomial expansion gives

\begin{equation}\label{ordr dens EWP}
\begin{array}[b]{ll}
f_{r:n}(y)&=\frac{1}{B(r,n-r+1)}\frac{\alpha\gamma \theta \beta
^{\gamma}}{(e ^{\theta}-1)^n} y^{\gamma -1}e ^
 {-(\beta y)^{\gamma}} (1 -e ^ {-(\beta y)^{\gamma}})^{\alpha -1} \medskip \\
&~~~\times \sum^{r-1}_{i=0}\sum^{n-r}_{j=0}(-1)^{r-i-j-1}
{r-1\choose{i}}{n -r\choose{j}} e^{(n-r-j)\theta} e^{(i+j)\theta
(1-e^{-(\beta y)^{\gamma}})^{\alpha}}.
\end{array}
\end{equation}
 Also the cdf of $Y_{r:n}$ is
 given by
\begin{equation}\label{ordr cdf EWP}
\begin{array}[b]{ll}
F_{r:n}(y)&=\sum^{n}_{k=r}{{n}\choose{k}}[F(y)]^{k}[1-F(y)]^{n-k}\medskip\\
&=‎\frac{1}{(e^{‎\theta‎} -1)^n}‎\sum^{n}_{k=r}
\sum^{k}_{i=0}\sum^{n-k}_{j=0}(-1)^{k-i-j} {{n}\choose{k}}
{k\choose{i}} {n-k \choose{j}} e^{(n-k-j)‎\theta‎}
e^{(i+j)‎\theta (1-e^{-(‎\beta
y‎)^{‎\gamma‎}})^{‎\alpha‎}‎} .
\end{array}
\end{equation}
Using the binomial expansion, after some calculations the $k$th
moment of the $r$th order statistic $Y_{r:n}$ is given by
\begin{equation}\label{Exp order EWP}
\begin{array}[b]{ll}
E(Y_{r:n} ^{k})&=\frac{ \alpha\theta
}{B(r,n-r+1) ‎\beta^{k‎}}\Gamma\left(\frac{k}{\gamma}+1\right)\sum^{n-r}_{i=0} ‎\frac{(-1)^i}{(e^{‎\theta} -1‎)^{r+i}}‎\sum^{r+i-1}_{j=0} (-1)^{r+i-j-1}{r+j-1\choose{j}}\medskip\\
&~~\times \sum^{\infty}_{l=0} ‎\frac{‎\theta ^{l}‎ (j+1)^{l}}{l!}‎\sum^{\infty}_{t=0} (-1)^{t}{‎\alpha (l+1) -1‎\choose{t}} (t+1)^{-(1+‎\frac{k}{‎\gamma‎}‎)} .
\end{array}
\end{equation}
The pdf and cdf of the smallest and biggest order statistics
$Y_{1:n}$ and $Y_{n:n}$ can be
 obtained using Eqs. (\ref{ordr dens EWP}) and (\ref{ordr cdf EWP}) for special cases $r=1$ and $n$.\\
 For the smallest order statistics $Y_{1:n}$, we have
\begin{equation*}
f_{1:n}(y)=\frac{n\alpha\gamma \theta \beta ^{\gamma}}{(e
^{\theta}-1)^n} y^{\gamma -1}e ^
 {-(\beta y)^{\gamma}} (1 -e ^ {-(\beta y)^{\gamma}})^{\alpha -1} e^{\theta(1 -e ^ {-(\beta
y)^{\gamma}})^{\alpha}}\left(e^{\theta}-e^{\theta(1 -e ^ {-(\beta
y)^{\gamma}})^{\alpha}}\right)^{n-1},
\end{equation*}
\begin{equation*}
F_{1:n}(y)=1-\frac{1}{(e^{\theta}-1)^{n}}\left(e^{\theta}-e^{\theta(1
-e ^ {-(\beta y)^{\gamma}})^{\alpha}}\right)^{n},
\end{equation*}
and for the biggest order statistics $Y_{n:n}$, we have
\begin{equation*}
f_{n:n}(y)=\frac{n\alpha\gamma \theta \beta ^{\gamma}}{(e
^{\theta}-1)^n} y^{\gamma -1}e ^
 {-(\beta y)^{\gamma}} (1 -e ^ {-(\beta y)^{\gamma}})^{\alpha -1} e^{\theta(1 -e ^ {-(\beta
y)^{\gamma}})^{\alpha}}\left(e^{\theta(1 -e ^ {-(\beta
y)^{\gamma}})^{\alpha}}-1\right)^{n-1},
\end{equation*}
\begin{equation*}
F_{n:n}(y)=\frac{1}{(e^{\theta}-1)^{n}}\left(e^{\theta(1 -e ^
{-(\beta y)^{\gamma}})^{\alpha}}-1\right)^{n},
\end{equation*}
The expectation of the smallest and biggest order statistics
$Y_{1:n}$ and $Y_{n:n}$ can be obtained by setting $r=1,n$ and $k=1$
in (\ref{Exp order EWP}).

\section{ Residual life function of the EWP distribution}
Given that a component survives up to time $t\geq0$, the residual
life is the period beyond $t$ until the time of failure and defined
by the conditional random variable $X-t|X > t$. The mean residual
life (MRL) function is an important function in survival analysis,
actuarial science, economics and other social sciences and
reliability for characterizing lifetime. Although the shape of the
failure rate function plays an important role in repair and
replacement strategies, the MRL function is more relevant as the
latter summarizes the entire residual life function, whereas the
former considers only the risk of instantaneous failure. In
reliability, it is well known that the MRL function and ratio of two
consecutive moments of residual life determine the distribution
uniquely (Gupta and Gupta, \cite{Gupta1983 }).

%Another important representation of the EWP is the mean residual
%life (MRL) function obtain by setting $r=1$ in (\ref{rresi}). MRL
MRL function as well as failure rate function is very important,
since each of them can be used to determine a unique corresponding
lifetime distribution. Lifetimes can exhibit IMRL (increasing MRL)
or DMRL (decreasing MRL). MRL functions that first decreases
(increases) and then increases (decreases) are usually called
bathtub-shaped (upside-down bathtub), BMRL (UMRL). The relationship
between the behaviors of the two functions of a distribution was
studied by many authors such as Ghitany \cite{Ghitany}, Mi
\cite{Mi}, Park \cite{Park } and Tang et al. \cite{Tang }.\\
The $r$th order moment of the residual life of the EWP distribution
is given by the general formula
\begin{equation*}
m_{r}(t)=E\left[(Y-t)^r|Y>t\right]=\frac{1}{S(t)}\int_{t}^{\infty}(y-t)^{r}f(y)dy,
\end{equation*}
where $S(t)=1-F(t)$, is the survival function.\newline In what seen
this onwards, we use the expressions

\begin{equation*}\label{incom int1}
\int_{t}^{\infty}x^{\gamma+s-1}e^{-(k+1)(\beta
x)^{\gamma}}dx=\frac{1}{\gamma\beta^{\gamma+s}}(k+1)^{-(1+\frac{s}{\gamma})}\Gamma^{(1+\frac{s}{\gamma})}\left((k+1)(\beta
t)^{\gamma}\right),
\end{equation*}
and
\begin{equation*}\label{incom int2}
\int_{0}^{t}x^{\gamma+s-1}e^{-(k+1)(\beta
x)^{\gamma}}dx=\frac{1}{\gamma\beta^{\gamma+s}}(k+1)^{-(1+\frac{s}{\gamma})}\Gamma_{(1+\frac{s}{\gamma})}\left((k+1)(\beta
t)^{\gamma}\right),
\end{equation*}
where $\Gamma^{s}(t)= \int_{t}^{\infty}x^{s-1}e^{-x}dx$ is the upper
incomplete gamma function and $\Gamma_{s}(t)=
\int_{0}^{t}x^{s-1}e^{-x}dx$ is the lower incomplete gamma function.
\\The $r$th order moment of the residual
life of the EWP distribution is given by
\begin{equation}\label{rresi}
\begin{array}[b]{ll}
m_{r}(t)&=\frac{‎\alpha ‎\theta ‎‎}{S(t)(e^{\theta}-1) }\sum_{i=0}^{r}(-1)^{r-i} {r\choose i}\beta^{-i} t^{r-i}\sum_{j=0}^{\infty}\sum_{k=0}^{\infty}(-1)^{k}\frac{\theta^{j}}{j!(k+1)^{1+\frac{i}{\gamma}}}\\
&~~\times{{\alpha(j+1)-1}\choose{k}}\Gamma^{\left(1+\frac{i}{\gamma}\right)}\left((k+1)(\beta
t)^{\gamma}\right),
\end{array}
\end{equation}
where $S(t)$ (the survival function of $Y$) is given in
(\ref{survive}). \\
For the EWP distribution the MRL function which is obtained by
setting $r=1$ in (\ref{rresi}), is given in the following theorem.

\begin{thm}
The MRL function of the EWP distribution with cdf (\ref{cdf EWP}) is
given by
\begin{equation*}\label{MRL EWG}
m_{1}(t)=\frac{1}{S(t)}\frac{\alpha\theta}{\beta(e^{\theta}-1)}\sum_{j=0}^{\infty}\sum_{k=0}^{\infty}(-1)^{k}\frac{\theta^{j}}{j!(k+1)^{1+\frac{1}{\gamma}}}
{{\alpha(j+1)-1}\choose{k}}\Gamma^{\left(1+\frac{1}{\gamma}\right)}((k+1)(\beta
t)^{\gamma})-t.
\end{equation*}
\end{thm}
The second moment of the residual life function of the EWP
distribution is
\begin{equation*}\label{r res}
\begin{array}[b]{ll}
m_{2}(t)&=\frac{1}{S(t)}\Big\{\frac{\alpha\theta}{\beta^{2}(e^{\theta}-1)}\Big[\sum_{j=0}^{\infty}\sum_{k=0}^{\infty}(-1)^{k}\frac{\theta^{j}}
{j!(k+1)^{1+\frac{2}{\gamma}}}{{\alpha(j+1)-1}\choose{k}}\Gamma^{\left(1+\frac{2}{\gamma}\right)}((k+1)(\beta
t)^{\gamma})\\
&~~-2\beta
t\sum_{j=0}^{\infty}\sum_{k=0}^{\infty}(-1)^{k}\frac{\theta^{j}}
{j!(k+1)^{1+\frac{1}{\gamma}}}{{\alpha(j+1)-1}\choose{k}}\Gamma^{\left(1+\frac{1}{\gamma}\right)}((k+1)(\beta
t)^{\gamma})\Big]\Big\}+t^2.
\end{array}
\end{equation*}
 The variance of the residual life function of the EWP distribution
can be obtained using $m_{1}(t)$ and $m_{2}(t)$.

 On the other hand,
we analogously discuss the reversed residual life and some of its
properties. The reversed residual life can be defined as the
conditional random variable $t-X|X \leq t$ which denotes the time
elapsed from the failure of a component given that its life is less
than or equal to $t$. This random variable may also be called the
inactivity time (or time since failure); for more details one may
see Kundu and Nanda \cite{Kundu2010} and Nanda et al. \cite{Nanda}.
Also, in reliability, the mean reversed residual life (MRRL) and
ratio of two consecutive moments of reversed residual life
characterize the distribution uniquely. Using (\ref{pdf EWP}) and
(\ref{survive}), the reversed failure (or reversed hazard) rate
function of the EWP is given by
\begin{equation*}
r(y)=\frac{f(y)}{F(y)}=\frac{\alpha\gamma\theta\beta ^{\gamma}
y^{\gamma -1} e^{-(\beta y)^{\gamma}}(1 -e^{-(\beta
y)^{\gamma}})^{\alpha -1} e^{\theta (1 -e^{-(\beta y)^{\gamma}}
)^{\alpha}}}{e^{\theta (1 -e^{-(\beta y)^{\gamma}} )^{\alpha}}-1}.
\end{equation*}
The $r$th moment of the reversed residual life function can be
obtained by the formula
\begin{equation*}
\mu_{r}(t)=E\left[(t-Y)^r|Y\leq
t\right]=\frac{1}{F(t)}\int_{0}^{t}(t-y)^{r}f(y)dy.
\end{equation*}
Hence,
\begin{equation}\label{rev residual}
\begin{array}[b]{ll}
\mu_{r}(t)&=\frac{1}{F(t)}\frac{\alpha\theta}{(e^{\theta}-1)}\sum_{i=0}^{r}(-1)^{i}{r\choose i}t^{r-i}\beta^{-i}\sum_{j=0}^{\infty}\sum_{k=0}^{\infty}(-1)^{k}\frac{\theta^{j}}{j!(k+1)^{1+\frac{i}{\gamma}}}\\
&~~\times{{\alpha(j+1)-1}\choose{k}}\Gamma_{\left(1+\frac{i}{\gamma}\right)}((k+1)(\beta
t)^{\gamma}).
\end{array}
\end{equation}
The mean and second moment of the reversed residual life of the EWP
distribution can be obtained by setting $r=1$ and $2$ in (\ref{rev
residual}). Also, using $\mu_{1}(t)$ and $\mu_{2}(t)$ one can obtain
the variance of the reversed residual life function of the EWP
distribution.

\section{ Probability weighted moments }
Probability weighted moments (PWMs) are expectations of certain
functions of a random variable defined when the ordinary moments of
the random variable exist. The PWMs method can generally be used for
estimating parameters of a distribution
whose inverse form cannot be expressed explicitly.\\
Estimates based on PWMs are often considered to be superior to
standard moment-based estimates. They are sometimes used when
maximum likelihood estimates are unavailable or difficult to
compute. They may also be used as starting values for maximum
likelihood estimates.

The PWMs method, which has been investigated by many researchers,
was originally proposed by Greenwood et al. \cite{Greenwood}. Since
then it has been used widely in practice and for research purposes.
Hosking et al. \cite{Hosking85} investigated the properties of
parameters estimated by the PWMs method for the generalized extreme
value (GEV) distribution using fairly long observed series, and they
gave a good summary of the PWMs method. Hosking \cite{Hosking86}
showed that the PWMs method is superior to the Maximum Likelihood
(ML) method in parameter estimations when the extreme value
distribution is used for longer return periods.

In this paper we calculate the PWMs of the EWP distribution since
they can be used for estimating the EWP parameters. For a random
variable with the pdf $f(.)$ and cdf $F(.)$, the PWMs function are
defined by
\begin{equation*}
\tau_{s,r}=E\left[X^{s}F(X)^{r}\right]=\int^{\infty}_{0}x^{s}(F(x))^{r}f(x)dx.
\end{equation*}
For the EWP distribution the PWMs are given by
\begin{equation*}
\begin{array}[b]{ll}
\tau_{s,r}&=\frac{\alpha \gamma\theta\beta^{\gamma}}{(e^{\theta}
-1)^{r+1}}\sum_{i=0}^{r}(-1)^{r-i}{r \choose
i}\int_{0}^{\infty}x^{\gamma+s-1}e^{-(\beta
x)^{\gamma}}(1-e^{-(\beta
x)^{\gamma}})^{\alpha-1}e^{\theta(i+1)(1-e^{-(\beta
x)^{\gamma}})^{\alpha}}dx\medskip\\
&=\frac{\alpha \gamma\theta\beta^{\gamma}}{(e^{\theta}
-1)^{r+1}}\sum_{i=0}^{r}(-1)^{r-i}{r \choose
i}\sum_{j=0}^{\infty}\frac{\theta^{j}(i+1)^{j}}{j!}\int_{0}^{\infty}x^{\gamma+s-1}e^{-(\beta
x)^{\gamma}}(1-e^{-(\beta
x)^{\gamma}})^{\alpha(j+1)-1}dx\medskip\\
&=\frac{\alpha \gamma\theta\beta^{\gamma}}{(e^{\theta}
-1)^{r+1}}\sum_{i=0}^{r}(-1)^{r-i}{r \choose
i}\sum_{j=0}^{\infty}\frac{\theta^{j}(i+1)^{j}}{j!}\sum_{k=0}^{\infty}(-1)^{k}{\alpha(j+1)-1
\choose k}\int_{0}^{\infty}x^{\gamma+s-1}e^{-(k+1)(\beta
x)^{\gamma}}dx.
\end{array}
\end{equation*}
Using formula $\int_{0}^{\infty}x^{\gamma+s-1}e^{-(k+1)(\beta
x)^{\gamma}}dx=\frac{\Gamma(1+\frac{s}{\gamma})}{\gamma\beta^{\gamma+s}(k+1)^{1+\frac{s}{\gamma}}}$,
we have
\begin{equation}\label{pwm}
\tau_{s,r}=\frac{\alpha \theta}{\beta^{s}(e^{\theta}
-1)^{r+1}}\sum_{i=0}^{r}(-1)^{r-i}{r \choose
i}\sum_{j=0}^{\infty}\frac{\theta^{j}(i+1)^{j}}{j!}\sum_{k=0}^{\infty}(-1)^{k}{\alpha(j+1)-1
\choose
k}\frac{\Gamma(1+\frac{s}{\gamma})}{(k+1)^{1+\frac{s}{\gamma}}}.
\end{equation}

\begin{rmk}
The $\textit{s}$th moment of the EWP distribution can be obtained by
putting $r=0$ in (\ref{pwm}). Therefore
\begin{equation*}
E(X^{s}) =\frac{\alpha \theta}{\beta^{s}(e^{\theta}
-1)}\sum_{j=0}^{\infty}\frac{\theta^{j}(i+1)^{j}}{j!}\sum_{k=0}^{\infty}(-1)^{k}{\alpha(j+1)-1
\choose
k}\frac{\Gamma(1+\frac{s}{\gamma})}{(k+1)^{1+\frac{s}{\gamma}}},
\end{equation*}
which is equal to Eq. (\ref{meankEWP2}), if $s$ is replaced by $k$.
Also, the mean and variance of the EWP distribution can be obtained
by setting $s=1$ and $2$ in above equation.
\end{rmk}

\section{ Mean deviations}

The amount of scatter in a population can be measured by the
totality of deviations from the mean and median. The mean deviation
from the mean is a robust statistic, being more resilient to
outliers in a data set than the standard deviation. In the standard
deviation, the distances from the mean are squared, so on average,
large deviations are weighted more heavily, and thus outliers can
heavily influence it. In the mean deviation from the mean, the
magnitude of the distances of a small number of outliers is
irrelevant. The mean deviation from the median is a measure of
statistical dispersion. It is a more robust estimator of scale than
the sample variance or standard deviation. It thus behaves better
with distributions without a mean or variance, such as the Cauchy
distribution.

For a random variable $X$ with pdf $f(x)$, cdf $F(x)$, mean $\mu$
and median $M$, the mean deviation from the mean and the mean
deviation from the median are defined by
$$\delta_{1}(X)=\int_{0}^{\infty}|x-\mu|f(x)dx=2\mu F(\mu)-2I(\mu),$$
and
$$\delta_{2}(X)=\int_{0}^{\infty}|x-M|f(x)dx=2M F(M)-M+\mu -
2I(M),$$ respectively, where $I(b)=\int^{b}_{0}xf(x)dx$

\begin{thm}
The Mean deviations function of the EWP distribution are
\begin{equation*}\label{MD EWp}
\begin{array}[b]{ll}
\delta_{1}(X)&=\frac{2}{e^{\theta}-1}\Big[\mu
\left(e^{\theta(1-e^{(-\beta
\mu)^{\gamma}})^{\alpha}}-1\right)-\frac{\alpha\theta}{\beta}\sum_{j=0}^{\infty}\sum_{k=0}^{\infty}(-1)^{k}\frac{\theta^{j}}{j!(k+1)^{1+\frac{1}{\gamma}}}
{{\alpha(j+1)-1}\choose{k}}\\&~~~\times\Gamma_{\left(1+\frac{1}{\gamma}\right)}((k+1)(\beta
\mu)^{\gamma}) \Big],
\end{array}
\end{equation*}
and
\begin{equation*}\label{MD EWp}
\delta_{2}(X)=\mu
-2\frac{\alpha\theta}{\beta(e^{\theta}-1)}\sum_{j=0}^{\infty}\sum_{k=0}^{\infty}(-1)^{k}\frac{\theta^{j}}{j!(k+1)^{1+\frac{1}{\gamma}}}
{{\alpha(j+1)-1}\choose{k}}\Gamma_{\left(1+\frac{1}{\gamma}\right)}((k+1)(\beta
M)^{\gamma}),
\end{equation*}
where $\mu$ is given in (\ref{meanEWP}) and $M$ can be obtained by
setting $p=.5$ in (\ref{p quantile}).
\end{thm}

\section{ Bonferroni and Lorenz curves}

Study of income inequality has gained a lot of importance over the
last many years. Lorenz curve and the associated Gini index are
undoubtedly the most popular indices of income inequality. However,
there are certain measures which despite possessing interesting
characteristics have not been used often for measuring inequality.
Bonferroni curve and scaled total time on test transform are two
such measures, which have the advantage of being represented
graphically in the unit square and can also be related to the Lorenz
curve and Gini ratio (Giorgi, \cite{Giorgi98}). These two measures
have some applications in reliability and life testing as well
(Giorgi and Crescenzi, \cite{Giorgi2001}). The Bonferroni and Lorenz
curves and Gini index have many applications not only in economics
to study income and poverty, but also in other fields like
reliability, medicine and insurance.

For a random variable $X$ with cdf $F(.)$, the Bonferroni curve is
given by
\begin{equation*}
B_{F}[F(x)]=\frac{1}{\mu F(x)}\int^{x}_{0}u f(u) du.
\end{equation*}
From the relationship between the Bonferroni curve and the mean
residual lifetime, the Bonferroni curve of the EWP distribution is
given by
\begin{equation*}\label{B_{F} EWp}
\begin{array}{ll}
B_{F}[F(x)]&=
\frac{\alpha\theta}{\beta\mu\left(e^{\theta(1-e^{(-\beta
x)^{\gamma}})^{\alpha}}-1\right)}\Big[
\sum_{j=0}^{\infty}\sum_{k=0}^{\infty}(-1)^{k}\frac{\theta^{j}}{j!(k+1)^{1+\frac{1}{\gamma}}}
{{\alpha(j+1)-1}\choose{k}}\\
&~~~\times\Gamma_{(1+\frac{1}{\gamma})}((k+1)(\beta
x)^{\gamma})\Big].
\end{array}
\end{equation*}
The Lorenz curve of the EWP distribution can be obtained via the
expression
\begin{equation*}
\begin{array}{ll}
L_{F} [F(x)] &=
B_{F}[F(x)]F(x)=\frac{\alpha\theta}{\beta\mu(e^{\theta}-1)}\sum_{j=0}^{\infty}\sum_{k=0}^{\infty}(-1)^{k}
\frac{\theta^{j}}{j!(k+1)^{1+\frac{1}{\gamma}}}
{{\alpha(j+1)-1}\choose{k}}\\
&~~~\times\Gamma_{(1+\frac{1}{\gamma})}((k+1)(\beta
x)^{\gamma})\Big],
\end{array}
 \end{equation*}
 where $\mu$ is the mean of the EWP distribution.\\
The scaled total time on test transform of a distribution function
$F$ is defined by
\begin{equation*}
S_{F}[F(t)]=\frac{1}{\mu}\int^{t}_{0}S(u)du.
\end{equation*}
If $F(t)$ denotes the cdf of the EWP distribution then
\begin{equation*}
S_{F}[F(t)]=\frac{1}{\mu(e^{\theta}-1)}\Big[te^{\theta}-\frac{1}{\gamma\beta}\sum^{\infty}_{j=0}\sum^{\infty}_{k=0}(-1)^{k}\frac{\theta^{j}
}{j!} {{\alpha j} \choose {k}}k^{-1/\gamma}\Gamma_{
(\frac{1}{\gamma})}(k(\beta t)^{\gamma})\Big].
\end{equation*}
The cumulative total time can be obtained by using formula
$C_F=\int_{0}^{1}S_{F}[F(t)]f(t)dt$ and the Gini index can be
derived from the relationship $G =1-C_{F}$.

\section{ Estimation and inference}
In this section, we discuss the estimation of the parameters of the
EWP distribution. Let $Y_{1},Y_{2},\cdots,Y_{n}$ be a random sample
with observed values $y_{1},y_{2},\cdots,y_{n}$ from EWP
distribution with parameters $\alpha,\beta,\gamma$ and $\theta$. Let
$\Theta=(\alpha,\beta,\gamma,\theta)^{T}$ be the parameter vector.
The total log-likelihood function is given by
\begin{equation*}
\begin{array}[b]{ll}
l_{n}\equiv l_{n}(y;\Theta)&= -n\log[e^{‎\theta‎}-1]+n[\log
‎‎\theta‎+\log ‎\alpha+ \log ‎\gamma +‎\gamma \log‎\beta‎ ]\medskip \\
&~~~+(‎\gamma -1‎)\sum^{n}_{i=1}\log
y_{i}-\sum^{n}_{i=1} (‎\beta y_i‎)^{\gamma} +(‎\alpha -1‎)‎\sum^{n}_{i=1} \log (1- e^{-(‎\beta y_i‎)^{‎\gamma‎}})\medskip \\
&~~~+ ‎\theta ‎\sum^{n}_{i=1} (1- e^{-(‎\beta
y_{i})^{‎\gamma‎}})^{‎\alpha‎}.
\end{array}
\end{equation*}
The associated score function is given by $U_{n}(\Theta)=(\partial
l_{n}/\partial \alpha,\partial l_{n}/\partial \beta, \partial
l_{n}/\partial \gamma,\partial l_{n}/\partial \theta)^{T}$, where

\begin{equation*}
\begin{array}{ll}
\frac{\partial l_{n}}{\partial
\alpha}&=‎\frac{n}{‎\alpha‎}+\sum^{n}_{i=1}  \log(1-
e^{-(‎\beta y_i‎)^{‎\gamma‎}})+ ‎\theta \sum^{n}_{i=1}
\log (1- e^{-(‎\beta y_i‎)^{‎\gamma‎}})‎
(1- e^{-(‎\beta y_i‎)^{‎\gamma‎}})^{‎\alpha‎}‎,\medskip \\
\frac{\partial l_{n}}{\partial \beta}&=‎\frac{n
‎\gamma‎}{‎\beta‎}- ‎\gamma ‎ \beta ^{‎\gamma
-1‎}\sum^{n}_{i=1} y_i^{‎\gamma‎} + (‎\alpha -1‎)‎\gamma
‎\beta ^{‎\gamma -1‎}\sum^{n}_{i=1}
‎\frac{y_i^{‎\gamma‎} e^{-(‎\beta y_i‎)^{‎\gamma‎}}}
{1-e^{-(‎\beta y_i‎)^{‎\gamma‎}}} \medskip \\
&~~+‎\theta ‎\gamma ‎\alpha‎ ‎\beta^{‎ \gamma
-1‎}‎‎‎‎‎‎‎‎‎\sum^{n}_{i=1} y_i^{‎\gamma‎}
e^{-(‎\beta y_i‎)^{‎\gamma‎}} (1-e^{-(‎\beta y_i‎)^
{‎\gamma‎}})^{‎\alpha -1‎},\medskip\\
\frac{\partial l_{n}}{\partial \gamma}&=‎\frac{n}{‎\gamma‎} +n
\log ‎\beta +\sum^{n}_{i=1} \log y _i - \sum^{n}_{i=1}
(‎\beta y_i‎)^{‎\gamma‎} \log (‎\beta y_i‎) \medskip \\
& ~~‎‎+ (‎\alpha -1‎) \sum^{n}_{i=1} ‎\frac{ (‎\beta
y_i‎)^{‎\gamma‎} \log (‎\beta y_i‎)‎‎ e^{-(‎\beta
y_i‎)^{‎\gamma‎}}}{1-e^{-(‎\beta y_i‎)^ {‎\gamma‎}}}+
‎\theta ‎\alpha\sum^{n}_{i=1} (‎\beta y_i‎)^{‎\gamma‎}
\log (‎\beta y_i‎) e^{-(‎\beta y_i‎)^{‎\gamma‎}}
(1-e^{-(‎\beta y_i‎)^
{‎\gamma‎}})^{‎\alpha -1‎},‎‎‎\medskip \\
\frac{\partial l_{n}}{\partial \theta}&= ‎\frac{n}{‎\theta‎}
+\sum^{n}_{i=1} (1-e^{-(‎\beta
y_{i}‎)^{‎\gamma‎}})^{‎\alpha ‎} -
‎\frac{n}{1-e^{-‎\theta‎}}‎‎.
\end{array}
\end{equation*}
The MLE of $\Theta $, say $\widehat{\Theta }$, is obtained by
solving the nonlinear system $U_n\left(\Theta \right)=\textbf{0}$.
The solution of this nonlinear system of equation has not a closed
form. For interval estimation and hypothesis tests on the model
parameters, we require the information matrix. The $4\times 4$
observed information matrix is
\[I_n\left(\Theta \right)=-\left[ \begin{array}{cccc}
I_{\alpha \alpha } & I_{\alpha \beta } & I_{\alpha \gamma } & I_{\alpha \theta }\\
I_{\alpha \beta } & I_{\beta \beta } & I_{\beta \gamma }& I_{\beta \theta } \\
I_{\alpha \gamma } & I_{\beta \gamma } & I_{\gamma \gamma}& I_{\gamma \theta } \\
I_{\alpha \theta } & I_{\beta \theta } & I_{\gamma \theta}& I_{\theta \theta } \\
\end{array} \right],\] whose elements are given in Appendix.

Applying the usual large sample approximation, MLE of $\Theta $ i.e.
$\widehat{\Theta }$ can be treated as being approximately
$N_4(\Theta ,{J_n(\Theta )}^{-1}{\mathbf )}$, where $J_n\left(\Theta
\right)=E\left[I_n\left(\Theta \right)\right]$. Under conditions
that are fulfilled for parameters in the interior of the parameter
space but not on the boundary, the asymptotic distribution of
$\sqrt{n}(\widehat{\Theta }{\rm -}\Theta {\rm )}$ is $N_4({\mathbf
0},{J(\Theta )}^{-1})$, where $J\left(\Theta \right)={\mathop{\lim
}_{n\to \infty } {n^{-1}I}_n(\Theta )\ }$ is the unit information
matrix. This asymptotic behavior remains valid if $J(\Theta )$ is
replaced by the average sample information matrix evaluated at
$\widehat{\Theta }$, say ${n^{-1}I}_n(\widehat{\Theta })$. The
estimated asymptotic multivariate normal $N_4(\Theta
,{I_n(\widehat{\Theta })}^{-1} )$ distribution of $\widehat{\Theta
}$ can be used to construct approximate confidence intervals for the
parameters and for the hazard rate and survival functions. A
$100(1-\gamma )$ asymptotic confidence interval for each parameter
${\Theta }_{{\rm r}}$ is given by
\[{ACI}_r=({\widehat{\Theta
}}_r-Z_{\frac{\gamma }{2}}\sqrt{{\hat{I}}^{rr}},{\widehat{\Theta
}}_r+Z_{\frac{\gamma }{2}}\sqrt{{\hat{I}}^{rr}}),\] where
${\hat{I}}^{rr}$ is the (\textit{r, r}) diagonal element of
${I_n(\widehat{\Theta })}^{-1}$ for $r=1,~2,~3,~4,$ and
$Z_{\frac{\gamma }{2}}$ is the quantile $1-\gamma /2$ of the
standard normal distribution.

\subsection{EM-algorithm}
The MLEs of the parameters $\alpha$, $\beta$, $\gamma$ and $\theta$
in previous section must be derived numerically. Newton-Raphson
algorithm is one of the standard methods to determine the MLEs of
the parameters. To employ the algorithm, second derivatives of the
log-likelihood are required for all iterations. The EM-algorithm is
a very powerful tool in handling the incomplete data problem
(Dempster et al., \cite{Dempster}; McLachlan and Krishnan,
\cite{McLachlan}). It is an iterative method by repeatedly replacing
the missing data with estimated values and updating the parameters.
It is especially useful if the complete data set is easy to analyze.

 Let the complete-data be $Y_{1},\cdots,Y_{n}$ with observed values $y_{1},\cdots,y_{n}$ and
the hypothetical random variable $Z_{1},\cdots,Z_{n}$. The joint
probability density function is such that the marginal density of
$Y_{1},\cdots,Y_{n}$ is the likelihood of interest. Then, we define
a hypothetical complete-data distribution for each
$(Y_{i},Z_{i})~~i=1,\cdots,n,$ with a joint probability density
function in the form
\begin{equation*}
g(y,z;\Theta)=f(y|z)f(z) = \frac{\theta^{z}}{z!(e^{\theta}- 1)}z
\alpha\gamma ‎\beta^{‎\gamma‎} y^{‎\gamma -1‎} e^{-(\beta
y)^{‎\gamma‎}} (1 -e^{-(\beta y )^{‎\gamma‎}})^{z ‎\alpha
-1‎},
\end{equation*}
where $\Theta=(\alpha,\beta,\gamma,\theta)$, $y>0$ and $z\in
\mathbb{N}$.\\
Under the formulation, the E-step of an EM cycle requires the
expectation of $(Z|Y;\Theta^{(r)})$ where
$\Theta^{(r)}=(\alpha^{(r)},\beta^{(r)},\gamma^{(r)},\theta^{(r)})$
is the current estimate of $\Theta$ (in the $r$th iteration).\\
The pdf of $Z$ given $Y$, say $g(z|y)$ is given by
\begin{equation*}
g(z|y)= ‎‎‎\frac{e^{ -‎\theta (1-e^{ -(‎\beta
y‎)^{‎\gamma‎}})^{‎\alpha‎}‎} [\theta (1-e^{ -(‎\beta
y)^{‎\gamma‎}})^{‎\alpha‎} ]^{(z-1)}}{(z-1)!}‎,
\end{equation*}
with the expectation
\begin{equation*}
\begin{array}[b]{ll}
E[Z|Y=y]&=1+‎\theta‎(1 - e^{-(‎\beta y‎)^{‎\gamma‎}}
)^{‎\alpha‎}.
\end{array}
\end{equation*}

The EM cycle is completed  with the M-step by using the maximum
likelihood estimation over $\Theta$, with the missing $Z$'s
replaced by  their conditional expectations given above.\\
The log-likelihood for the complete-data is
\begin{equation*}
\begin{array}{ll}
l^{*}_{n}(y_{1},\cdots,y_{n};z_{1},\cdots,z;\Theta)&\propto
\sum^{n}_{i=1} z_i \log \theta -
n\log(e^{\theta}-1)+n\log \alpha +n \log \gamma +n\gamma \log \beta \medskip \\
& +(\gamma-1)\sum^{n}_{i=1} \log y_i -\sum^{n}_{i=1} (\beta y_i
)^{\gamma}+\sum^{n}_{i=1} (z_i\alpha -1) \log(1-e^{-{(\beta y_i
)}^{\gamma}}).
\end{array}
\end{equation*}

The components of the score function $U^{*}_{n}(\Theta)=
(\frac{\partial l^{*}_{n}}{\partial \alpha},\frac{\partial l^{*}_{n}
}{\partial \beta},\frac{\partial l^{*}_{n}}{\partial \gamma
},\frac{\partial l^{*}_{n}}{\partial \theta})^{T}$ are given by
\begin{equation*}
\begin{array}{ll}
\frac{\partial
l^{*}_{n}}{\partial\alpha}&=\frac{n}{\alpha}+ \sum^{n}_{i=1}  z_i \log (1 -e^{-(\beta y_i )^{\gamma}}),\medskip\\
\frac{\partial l^{*}_{n}}{\partial\beta}&=\frac{n\gamma}{\beta}
-\gamma\beta^{\gamma
-1}\sum^{n}_{i=1}y_i^{\gamma}+\gamma\beta^{\gamma -1}\sum^{n}_{i=1}
(z_i\alpha -1)[\frac{y_i^{\gamma}e^{-(\beta y_i )^{\gamma}}}{1- e^{-(\beta y_i )^{\gamma}}}],\medskip\\
\frac{\partial
l^{*}_{n}}{\partial\gamma}&=‎\frac{n}{‎\gamma‎}‎ +n
\log‎\beta +‎ \sum^{n}_{i=1}\log y_i - \sum^{n}_{i=1} (‎ \beta
y_i‎)^{‎\gamma‎} \log(‎\beta y_i‎)+  \sum^{n}_{i=1}(z_i
‎\alpha -1‎) ‎\frac{(\beta y_i )^{‎\gamma‎}
\log(\beta y_i ) e^{-( ‎\beta‎y_i)^{‎\gamma‎}}}{1-e^{-(\beta y_i )^{‎\gamma‎}}}‎,\medskip\\
\frac{\partial l^{*}_{n}}{\partial\theta}&=‎\frac{1}{‎\theta‎}
\sum^{n}_{i=1} z_i -‎\frac{n}{1-e^{-‎\theta‎}}‎‎.
\end{array}
\end{equation*}
From a nonlinear system of equations $U^{*}_{n}(\Theta)=\textbf{0}$,
we obtain the iterative procedure of the EM-algorithm as
\begin{equation*}
\begin{array}{l}
\hat{\alpha}^{(t+1)}=\frac{-n}{\sum^{n}_{i=1} z_i^{(t)}\log (1- e^{-( \hat{\beta}^{(t)}y_{i})^{\hat {\gamma}^{(t)}}})},\medskip\\
\frac{n \hat {\gamma}^{(t)}}{\hat{\beta} ^{(t+1)}} -\hat{\gamma}
^{(t)}(\hat{\beta}^{(t+1)})^{\hat {\gamma}^{(t)}-1}\sum^{n}_{i=1}
y_i^{\hat{\gamma}^{(t)}} +\hat{\gamma}
^{(t)}(\hat{\beta}^{(t+1)})^{\hat {\gamma}^{(t)}-1}\sum^{n}_{i=1}
(z_{i}^{(t)} \hat{\alpha}^{(t)} -1) \frac{y_i^{\hat{\gamma}^{(t)}}
e^{-( \hat {\beta}^{(t+1)}y_i)^{\hat{\gamma}^{(t)}}}}{1-e^{ -(
\hat{\beta}^{(t+1)}y_i)^{\hat{\gamma}^{(t)}}}}
=0,\medskip\\
\frac{n}{\hat{\gamma}^{(t+1)}}+n\log \hat{\beta}^{(t)}+
\sum^{n}_{i=1}\log y_{i}-\sum^{n}_{i=1}\log(\hat{\beta}^{(t)} y_i)(\hat{\beta}^{(t)}y_i)^{\hat{\gamma}^{(t+1)}} \medskip \\
+\sum^{n}_{i=1}(z_i^{(t)} \hat {\alpha} ^{(t)}-1) \frac{( \hat{\beta
}^{(t)}y_i)^{ \hat{\gamma} ^{(t+1)}} \log (\hat{\beta }^{(t)}y_i)
e^{-( \hat {\beta} ^{( t)}y_i)^{\hat{\gamma}
^{(t+1)}}}}{1-e^{-(\hat{\beta}^{(t)}y_i)^{\hat{\gamma}^{(t+1)}} }}=0,\medskip\\
\frac{\sum^{n}_{i=1} z_i^{(t)}}{\hat{\theta }^{(t+1)}}
-\frac{n}{1-e^{-\hat{\theta}^{(t+1)}}}=0,
\end{array}
\end{equation*}
where $\hat\alpha^{(t+1)}$, $\hat\beta^{(t+1)}$ and
$\hat\theta^{(t+1)}$ are found numerically. Hence, for
$i=1,\cdots,n$, we have that
\begin{equation*}
z^{(t)}_{i}=1+\hat{\theta}^{(t)} (1-e^{-( \hat{\beta}^{ (t)}y_i)^{
\hat {\gamma }^{(t)}}})^{\hat{\alpha} ^{(t)}}.
\end{equation*}

\section{Sub-models of the EWP distribution }
The EWP distribution contains some sub-models for the special values
of $\alpha$, $\beta$ and $\gamma$. Some of these distributions are
discussed here in details.

\subsection{The CWP distribution}
The CWP distribution is a special case of EWP distribution for
$\alpha=1$. Our approach here is complementary to that of Morais and
Barreto-Souza \cite{Morais } in introducing the Weibull-Poisson (WP)
distribution. The pdf, cdf and hazard rate function of the CWP
distribution are given, respectively by
\begin{equation*}\label{pdf CWP}
f(y)=\frac{\gamma\theta \beta ^{\gamma}}{e^{\theta}-1}y^{\gamma -1}
e^{-(\beta y)^{\gamma}} e^{\theta(1-e^{-(\beta y)^{\gamma}})},
\end{equation*}
\begin{equation*}\label{cdf CWP}
F(y)=\frac{e^{\theta (1-e^{-(\beta y)^{\gamma}})} -1}{e^{\theta}
-1},
\end{equation*}
and
\begin{equation*}\label{hazard CWP}
h(y)=‎‎\frac{‎\gamma‎\theta ‎\beta
^{‎\gamma‎}‎‎‎ y^{‎\gamma -1‎} e^{-(‎\beta
y‎)^{‎\gamma‎}} e^{‎\theta(1-e^{-(‎\beta
y‎)^{‎\gamma‎}})‎}}{e^{‎\theta‎} -e^{‎\theta
(1-e^{-(‎\beta y‎)^{‎\gamma‎}})‎}}‎‎‎.
\end{equation*}
According to (\ref{meanEWP}) and (\ref{varEWP}), the mean and
variance of the CWP distribution are given, respectively by
\begin{equation}\label{mean CWP}
E(Y)=\frac{\theta\Gamma (1+\frac{1}{\gamma}
)}{\beta(e^{\theta}-1)}\sum^{\infty}_{n=1}\sum^{n-1}_{j=0}(-
1)^{j} \frac{\theta^{n-1}}{(n-1)!}{n - 1 \choose j}
(j+1)^{-(1+\frac{1}{\gamma})} ,
\end{equation}
and
\begin{equation*}
%\begin{array}[b]{ll}
Var(Y)=\frac{\theta\Gamma (1+\frac{2}{\gamma} )}{\beta^{
2}(e^{\theta}-1)}\sum^{\infty}_{n=1}\sum^{n-1}_{j=0}(- 1)^{j}
\frac{\theta^{n-1}}{(n-1)!}{n - 1 \choose j}
(j+1)^{-(1+\frac{2}{\gamma})}-E^2(Y),
%\end{array}
\end{equation*}
where $E(Y)$ is given by Eq. (\ref{mean CWP}). One can obtain the
Weibull distribution from the CWP distribution by taking $\theta$ to
be close to zero.

\subsection{The GEP distribution  }

GEP distribution is a special case of the EWP distribution for
$\gamma$=1. This distribution is introduced and analyzed in details
by Mahmoudi and jafari \cite{Mahmoudi2011a }. The pdf, cdf and
hazard rate function of the GEP distribution are given, by
\begin{equation*}\label{pdf GEP}
f(y)=\frac{\alpha\beta\theta }{e^{\theta} -1} e^{-\beta y}
(1-e^{-(\beta y)}) ^{\alpha-1}e^{\theta(1-e^{-\beta y})^{\alpha}},
\end{equation*}
\begin{equation*}\label{cdf GEP}
F(y)=\frac{e^{\theta (1-e^{-\beta y})^{\alpha}} -1}{e^{\theta} -1},
\end{equation*}
and
\begin{equation*}\label{hazard GEP}
h(y)=\frac{\alpha\beta\theta e^{-\beta y} (1-e^{-\beta y})
^{\alpha-1} e^{\theta(1-e^{-\beta y})^{\alpha}}}{e^{\theta}
-e^{\theta (1-e^{-\beta y})^{\alpha}}},
\end{equation*}
respectively. The mean and variance of GEP distribution are
\begin{equation}\label{meanGEP}
\begin{array}[b]{ll}
  E(Y) & =\frac{\alpha\theta}{\beta(e^{\theta} -1)}
\sum^{\infty}_{n=1}\sum^{\infty}_{j=0}(-1)^{j}\frac{\theta^{n-1}}{(n-1)!}\frac{{n\alpha-1
\choose{j}}}{(j+1)^2}\\ \medskip
   & =\frac{\theta}{\beta(e^{\theta} -1)}\sum^{\infty
}_{n=1}\frac{\theta ^{n-1}}{n!}\psi(n\alpha +1)-\frac{1}{\beta }\psi
(1),
\end{array}
\end{equation}
and
\begin{equation*}
\begin{array}{ll}
  Var(Y) & =\frac{2\alpha\theta}{\beta^{
2}(e^{\theta}-1)}\sum^{\infty}_{n=1}\sum^{\infty}_{j=0}(- 1)^{j}
\frac{\theta^{n-1}}{(n-1)!}\frac{{n\alpha - 1 \choose j}}
{(j+1)^{3}}-E^2(Y)\\ \medskip
   & =\frac{1}{{\beta }^2}{\psi
}'(1)-\frac{\theta}{\beta ^{2}(e^{\theta }-1)}\sum^{\infty
}_{n=1}{\frac{{\theta }^{n-1}}{n!}\left(\psi '(n\alpha
+1)+{\left(\psi (n\alpha +1)-\psi (1)\right)}^2\right)}-E^2(Y),
\end{array}
\end{equation*}
where $E(Y)$ is given in (\ref{meanGEP}), $\psi (.)$ is the digamma
function and $\psi'(.)$ is its derivative.

\subsection{The CEP distribution }

The CEP distribution is another special case of EWP distribution
obtained by replacing $\alpha=\gamma=1$, in EWP distribution. Cancho
et al. \cite{Cancho } introduced this distribution as the name
Poisson-exponential distribution. This distribution has an
increasing failure rate and arises on a latent complementary risk
problem base. The pdf, cdf and hazard rate function of the CEP
distribution are given, respectively by
\begin{equation*}\label{pdf CEP}
f(y)=\frac{\theta \beta }{e^{\theta} -1} e^{-\beta y}
e^{\theta(1-e^{-\beta y})},
\end{equation*}
\begin{equation*}\label{cdf CEP}
F(y)=\frac{e^{\theta (1-e^{-\beta y})} -1}{e^{\theta} -1},
\end{equation*}
and
\begin{equation*}\label{hazard CEP}
h(y)=\frac{\theta \beta e^{-\beta y} e^{\theta(1-e^{-\beta
y})}}{e^{\theta} -e^{\theta (1-e^{-\beta y})}}.
\end{equation*}
The mean and variance of the CEP distribution are given by
\begin{equation}\label{mean GEP}
\begin{array}[b]{ll}
  E(Y) & =\frac{\theta}{\beta(e^{\theta} -1)}
\sum^{\infty}_{n=1}\sum^{n-1}_{j=0}(-1)^{j}\frac{\theta^{n-1}}{(n-1)!}\frac{{n-1
\choose{j}}}{(j+1)^2}\\ \medskip
   & =\frac{\theta}{\beta(e^{\theta} -1)}\sum^{\infty
}_{n=1}\frac{\theta ^{n-1}}{n!}\psi(n+1)-\frac{1}{\beta }\psi (1),
\end{array}
\end{equation}
and
\begin{equation*}
\begin{array}{ll}
  Var(Y) & =\frac{2\theta}{\beta^{
2}(e^{\theta}-1)}\sum^{\infty}_{n=1}\sum^{n-1}_{j=0}(- 1)^{j}
\frac{\theta^{n-1}}{(n-1)!}\frac{{n - 1 \choose j}}
{(j+1)^{3}}-E^2(Y)\\ \medskip
   & =\frac{1}{{\beta }^2}{\psi
}'(1)-\frac{\theta}{\beta ^{2}(e^{\theta }-1)}\sum^{\infty
}_{n=1}{\frac{{\theta }^{n-1}}{n!}\left(\psi '(n+1)+{\left(\psi
(n+1)-\psi (1)\right)}^2\right)}-E^2(Y),
\end{array}
\end{equation*}
respectively, where $E(Y)$ is given in Eq. (\ref{mean GEP})

\section{Applications of the EWP distribution}

To show the superiority of the EWP distribution, we compare the
results of fitting this distribution to some of theirs sub-models
such as EW, GE and Weibull distributions, using two real data sets.
The required numerical evaluations are implemented using the SAS and
R softwares.
 The empirical scaled TTT transform (Aarset, \cite{Aarset}) and Kaplan-Meier curve can be used to identify the shape of the
hazard function.
%The scaled TTT transform is convex (concave) if the
%hazard rate is decreasing ( increasing), and for bathtub (unimodal)
%hazard rates, the scaled TTT transform is first convex (concave) and
%then concave (convex).

The first data consists of 63 record of the strengths of 1.5 cm
glass fibres, measured at the National Physical Laboratory, England.
Barreto-Souza et al. \cite{Barreto-Souza2010 } used this data to fit
the beta-generalized exponential (BGE) distribution.
\begin{figure}[]
\centering
\includegraphics[scale=0.40]{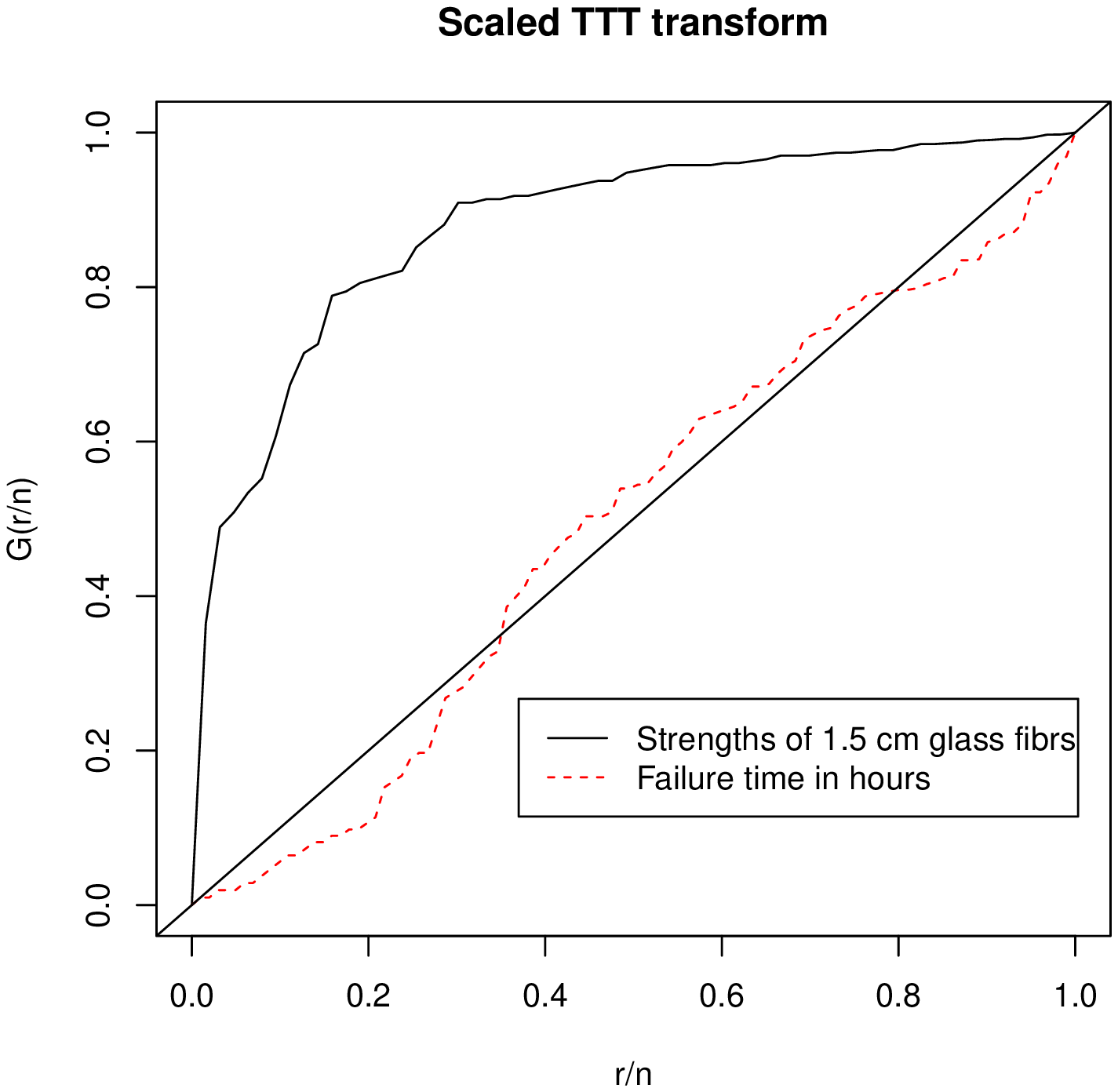}
\includegraphics[scale=0.40]{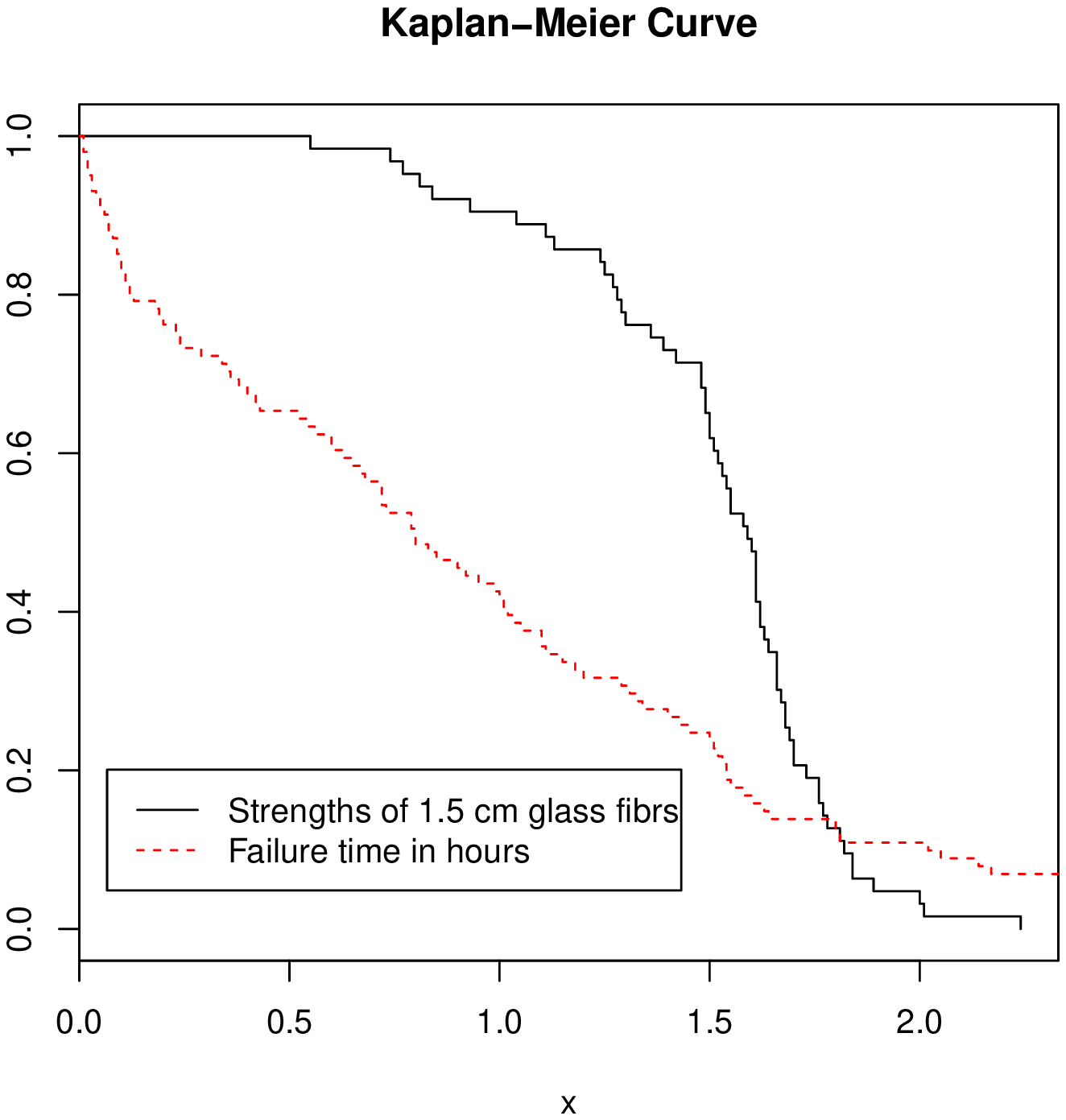}
\caption[]{TTT plots and Kaplan-Meier curves of strengths of 1.5 cm
glass fibres and the failure times in hours data.}
\end{figure}
The TTT plot and Kaplan-Meier curve for the first data in Fig. 3
shows an increasing hazard rate function and, therefore, indicates
that appropriateness of the EWP distribution to fit this data. Table
1 lists the MLEs with the standard deviations of the parameters, the
values of K-S (Kolmogorov-Smirnov) statistic with its respective
\textit{p}-value, -2log(L), AIC (Akaike Information Criterion), AD
(Anderson-Darling statistic) and CM (Cram\'{e}r-von Mises statistic)
for the first data. These values show that the EWP distribution
provide a better fit than the EW, GE and Weibull for fitting the
first data.

We apply the AD and CM statistics, in order to verify which
 distribution fits better to this data. The AD and CM test statistics are described in details in
Chen and Balakrishnan \cite{Chen}. In general, the smaller the
values of AD and CM, the better the fit to the data. According to
these statistics in Table 1, the EWP distribution fit the first data
set better than the others.
\begin{table}[htp!]
\centering \caption{MLEs(stds.), K-S statistics, \textit{p}-values,
$-2\log(L)$ and AIC for the strengths of 1.5 cm glass fibres.}
\begin{small}
\begin{tabular}{|l|lcccccc|}
\hline
Dist.& MLEs(stds.) & K-S  & \textit{p}-value &$-2\log(L)$& AIC& AD& CM  \\
\hline EWP & $\begin{array}{l}
 \hat{\alpha}=0.578 (0.3424),
\hat{\beta}=0.647 (0.0508)\\
\hat{\gamma}=5.502 (1.395), \hat{\theta}=2.782 (1.568)
\end{array}$
&0.1154 &0.3731  &26.0& 34.0&0.682&0.195 \\
EW & $\begin{array}{l}
 \hat{\alpha}=0.671(0.2489),
\hat{\beta}=0.582(0.0292)\\
\hat{\gamma}=7.285(1.707)
\end{array}$
&0.1901 &0.0211  &29.4& 35.4&1.086&0.279 \\
%\hline
GE &~ $\hat{\alpha}$=31.349(9.519), $\hat{\beta}$=2.612(0.2380)&0.2291  &0.0027&62.8  &66.8&4.340&0.881  \\
%\hline
Weibull &~  $\hat{\beta}$=0.614(0.0140), $\hat{\gamma}$=5.781(0.5761)&0.1589  &0.0830&30.4  &34.4 &3.135&0.297\\
\hline
\end{tabular}
\end{small}
\end{table}

Using the likelihood ratio (LR) test, we test the null hypothesis
H0: EW versus the alternative hypothesis H1: EWP, or equivalently,
H0: $\theta=0$ versus H1: $\theta\neq 0$. The value of the LR test
statistic and the corresponding \textit{p}-value are 3.4 and 0.0651,
respectively. Therefore, the null hypothesis (EW model) is rejected
in favor of the alternative hypothesis (EWP model) for a
significance level $>$ 0.0651. For test the null hypothesis H0: GE
versus the alternative hypothesis H1: EWP, or equivalently, H0:
$(\gamma,\theta)=(1,0)$ versus H1: $(\gamma,\theta)\neq(1,0)$, the
value of the LR test statistic is 36.8 (\textit{p}-value = 1e-08),
which includes that the null hypothesis (GE model) is rejected in
favor of the alternative hypothesis (EWP model) for any significance
level. We also test the null hypothesis H0: Weibull versus the
alternative hypothesis H1: EWP, or equivalently, H0:
$(\alpha,\theta)=(1,0)$ versus H1: $(\alpha,\theta)\neq(1,0)$. The
value of the LR test statistic is 4.4 (\textit{p}-value = 0.1108),
which includes that the null hypothesis (Weibull model) is rejected
in favor of the alternative hypothesis (EWP model) for a
significance level $>$ 0.1108, and for any significance level $<$
0.1108, the null hypothesis is not rejected but the values of AD and
CM in Table 1 show that the EWP distribution gives the better fit to
the first data set than the Weibull distribution.

As a second application, we consider the data show the
stress-rupture life of kevlar 49/epoxy strands which were subjected
to constant sustained pressure at the 90 stress level until all had
failed. The failure times in hours are shown in Andrews and Herzberg
\cite{Andrews } and Barlow et al. \cite{Barlow}. The TTT plot and
Kaplan-Meier curve for this data in Fig. 3 shows bathtub-shaped
hazard rate function and, therefore, indicates that appropriateness
of the EWP distribution to fit this data. The MLEs with the standard
deviations of the parameters, the values of K-S statistic,
\textit{p}-value, -2log(L), AIC, AD and CM are listed in Table 2.
From these values, we note that the EWP model is better than the EW,
GE and Weibull distributions in terms of fitting to this data.

\begin{table}[htp!]
\centering \caption{MLEs(stds.), K-S statistics, \textit{p}-values,
$-2\log(L)$ and AIC for the strengths of 1.5 cm glass fibres.}
\begin{small}
\begin{tabular}{|l|lcccccc|}
\hline
Dist.& MLEs(stds.) & K-S  & \textit{p}-value &$-2\log(L)$& AIC& AD& CM  \\
\hline EWP & $\begin{array}{l}
 \hat{\alpha}=0.859 (0.3681),
\hat{\beta}=1.303 (0.7398)\\
\hat{\gamma}=0.8717 (0.2409), \hat{\theta}=1.266 (1.2007)
\end{array}$
&0.0725 &0.6640  &204.6& 212.6&0.841&0.218\\
EW & $\begin{array}{l}
 \hat{\alpha}=0.793(0.2870),
\hat{\beta}=0.821(0.2651)\\
\hat{\gamma}=1.060(0.2399)
\end{array}$
&0.0844 &0.4677  &205.6& 211.6&0.955&0.247 \\
%\hline
GE &~ $\hat{\alpha}$=0.866(0.1098), $\hat{\beta}$=0.888(0.1201)&0.0887  &0.4041&205.6  &209.6&1.021&0.263  \\
%\hline
Weibull &~  $\hat{\beta}$=1.010(0.1141), $\hat{\gamma}$=0.926(0.0726)&0.0907  &0.3777&206.0  &210.0 &1.224&0.2789\\
\hline
\end{tabular}
\end{small}
\end{table}

Plots of the estimated pdf, cdf and survival function of the EWP,
EW, GE and Weibull models fitted to the data sets corresponding to
Tables 1 and 2, respectively, are given in Fig. 4. These plots
suggest that the EWP distribution is superior to the EW, GE and
Weibull distributions in fitting these two data sets.

\section{ Conclusion}

A new four-parameter distribution called the EWP distribution is
introduced. This distribution is a generalization of the EW
distribution and contains several lifetime sub-models such as: GEP,
CWP, CEP, ERP and RP. A characteristic of the EWP distribution is
that its failure rate function can be decreasing, increasing,
bathtub-shaped and unimodal depending on its parameter values.
Several properties of the new distribution such as its probability
density function, its reliability and failure rate functions,
quantiles and moments is obtained. The maximum likelihood estimation
procedure via a EM-algorithm is presented. Fitting the EWP model to
two real data sets indicates the flexibility and capacity of the
proposed distribution in data modeling. In view of the density and
failure rate function shapes, it seems that the proposed model can
be considered as a suitable candidate model in reliability analysis,
biological systems, data modeling, and related fields.

\section*{Acknowledgement}
The authors would like to sincerely thank the Editor-in-Chief,
Associate Editor, and referees for carefully reading the paper and
for their useful comments. The authors are also indebted to Yazd
University for supporting this research.

\begin{figure}
\centering
\includegraphics[scale=0.4]{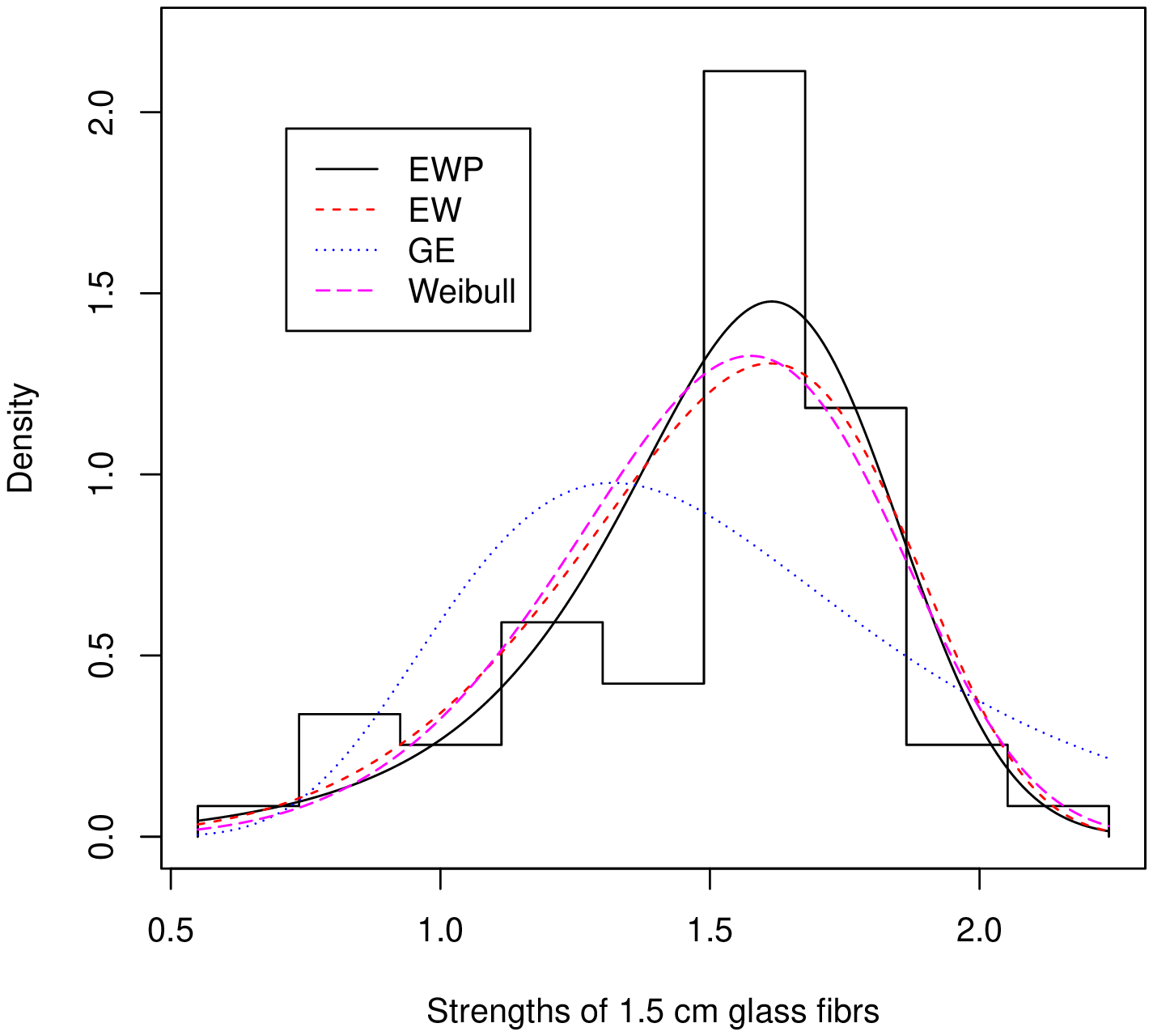}
\includegraphics[scale=0.4]{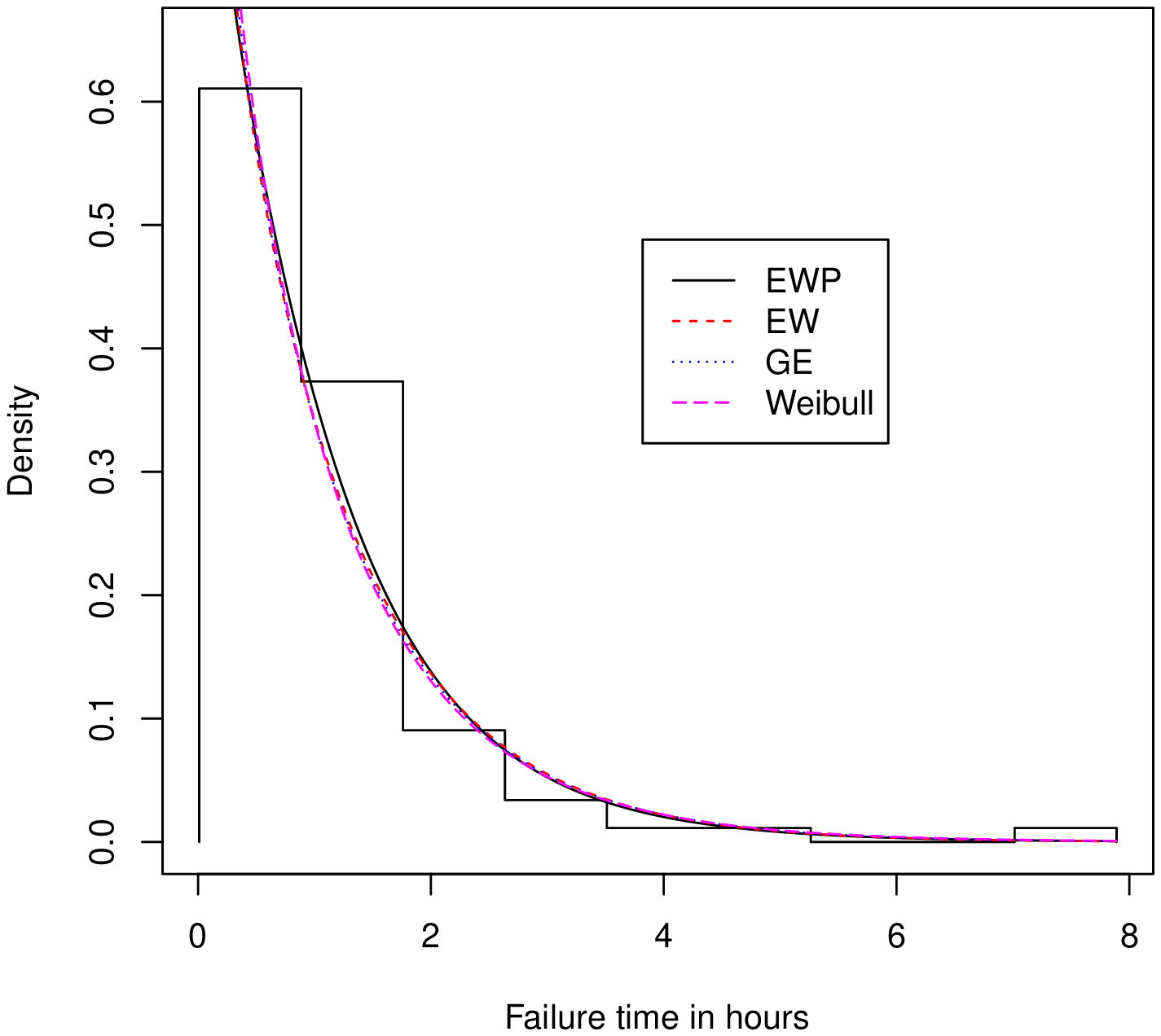}
\includegraphics[scale=0.40]{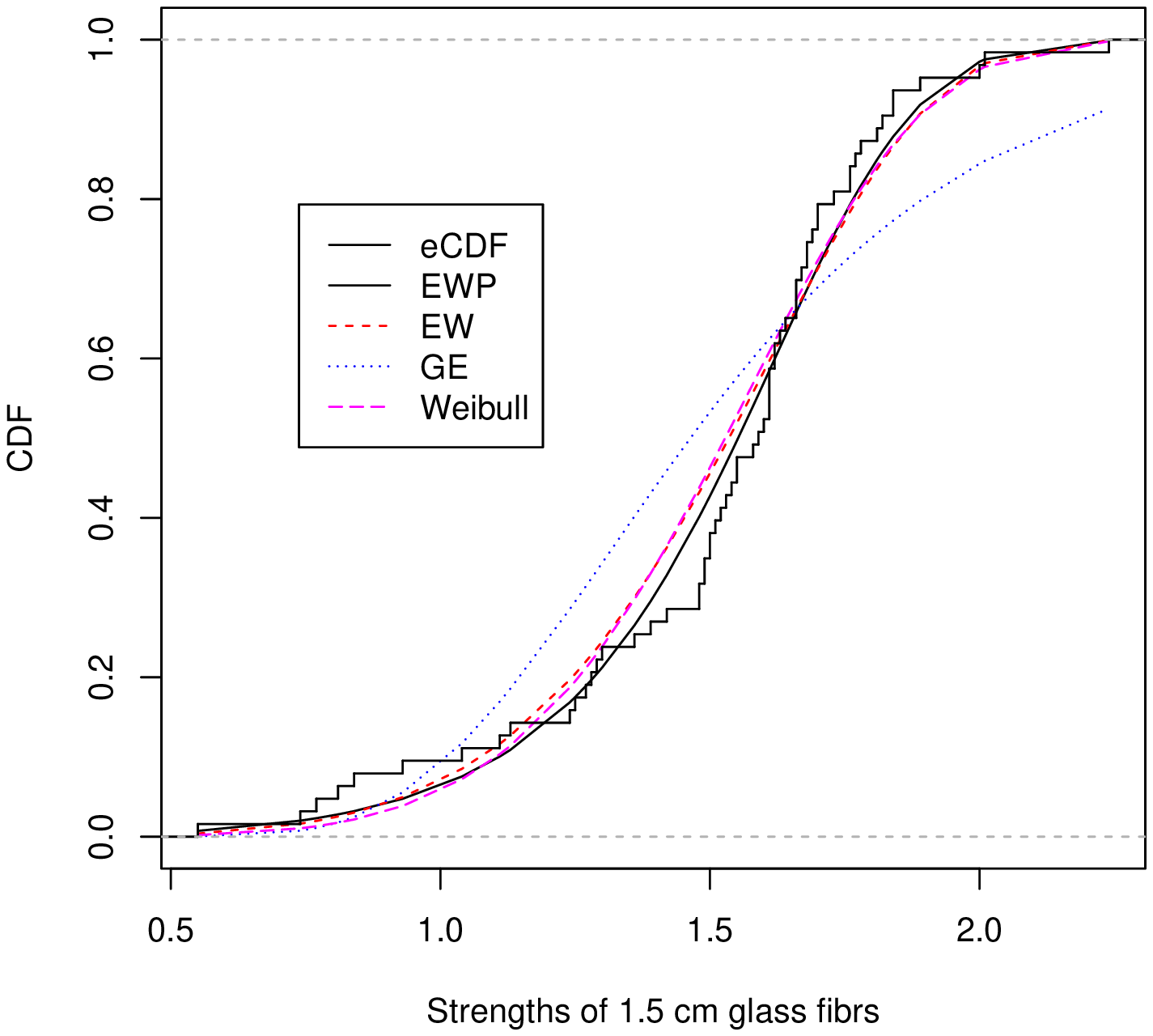}
\includegraphics[scale=0.4]{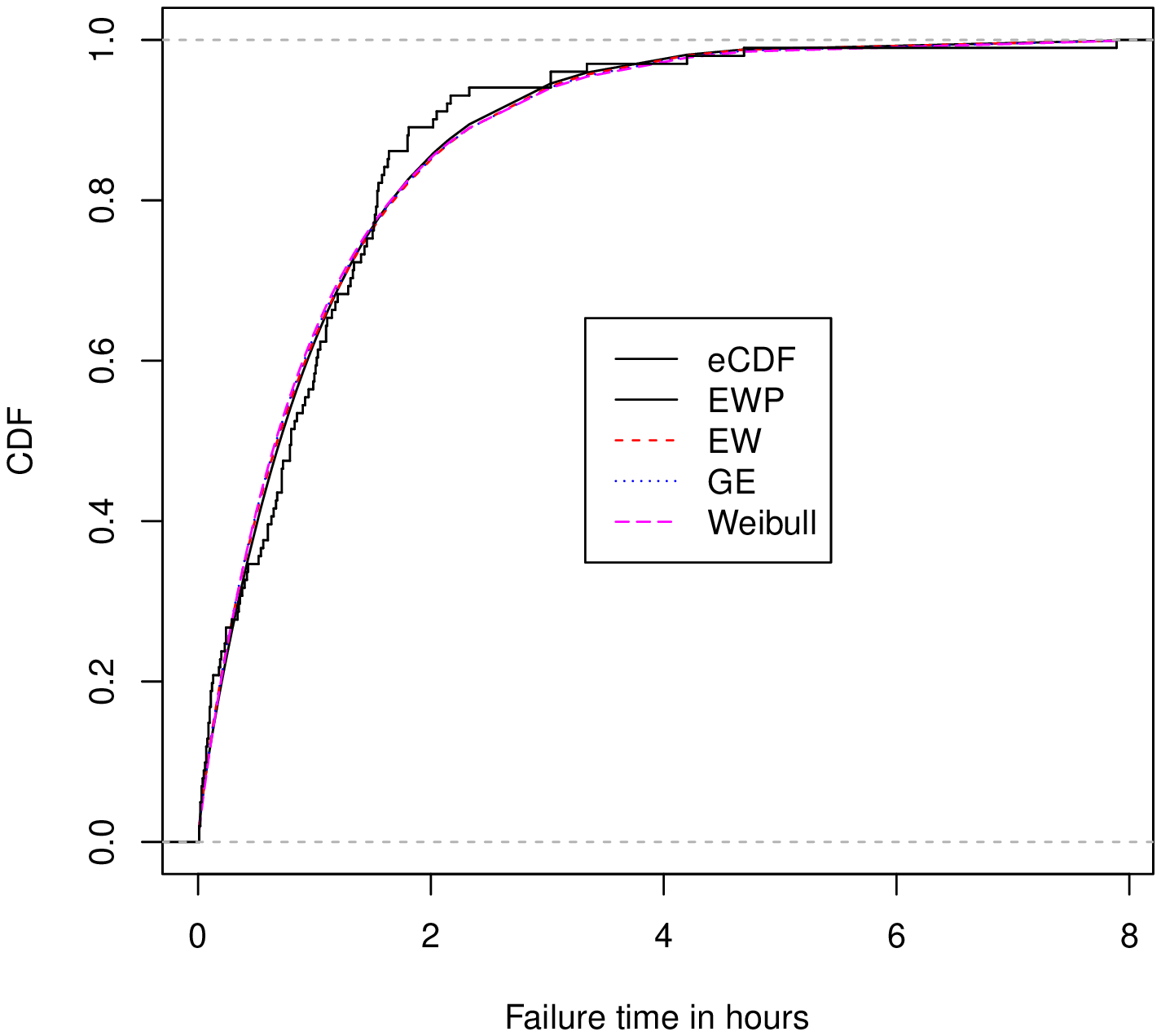}
\caption[]{Fitted pdf and cdf of the EWP, EW, GE and Weibull
distributions for the data sets corresponding to Tables 1 and 2,
respectively.}
\end{figure}

\newpage
\section*{Appendix}

The elements of $4\times 4$ observed information matrix
$I_n\left(\Theta \right)$ are given by

\begin{equation*}
\begin{array}{ll}
  I_{\alpha\alpha} &= -\frac{n}{\alpha ^2}+\theta  \sum _{i=1}^n
\left(1-e^{-\left(\beta  y_i\right)^{\gamma }}\right)^{\alpha }
\log\left(1-e^{-\left(\beta y_i\right){}^{\gamma }}\right)^2,\bigskip \\
  I_{\alpha\beta} & =\gamma \sum _{i=1}^n y_i \frac{e^{-\left(\beta  y_i\right)^{\gamma }}\left(\beta  y_i\right)^
  {\gamma-1 }}{1-e^{-\left(\beta  y_i\right)^{\gamma }}}+\theta\gamma\sum _{i=1}^n y_i e^{-\left(\beta  y_i\right)^{\gamma }}
  \left(1-e^{-\left(\beta  y_i\right)^{\gamma }}\right)^{\alpha-1 }\left(\beta  y_i\right)^{\gamma-1 }\medskip \\
  &~~+\theta\alpha\gamma  \sum _{i=1}^n y_i e^{-\left(\beta  y_i\right)^{\gamma }} \left(1-e^{-\left(\beta  y_i\right)^{\gamma }}\right){}^{\alpha-1 }
  \log\left(1-e^{-\left(\beta  y_i\right)^{\gamma }}\right) \left(\beta
  y_i\right)^{\gamma-1},\bigskip \\
  I_{\alpha\theta} & =\sum _{i=1}^n \left(1-e^{-\left(\beta
y_i\right){}^{\gamma }}\right){}^{\alpha }
\log\left(1-e^{-\left(\beta y_i\right){}^{\gamma }}\right),\bigskip \\
  I_{\alpha\gamma} & =\sum _{i=1}^n \frac{e^{-\left(\beta  y_i\right){}^{\gamma }} \log\left(\beta
  y_i\right)
  \left(\beta  y_i\right)^{\gamma }}{1-e^{-\left(\beta  y_i\right)^{\gamma }}}+\theta  \sum _{i=1}^n
  e^{-\left(\beta  y_i\right)^{\gamma }} \left(1-e^{-\left(\beta  y_i\right)^{\gamma }}\right){}^{\alpha -1}
  \log\left(\beta  y_i\right) \left(\beta  y_i\right){}^{\gamma }\medskip \\
   & ~~+\alpha\theta\sum _{i=1}^n e^{-\left(\beta  y_i\right)^{\gamma }}\left(1-e^{-\left(\beta  y_i\right)^{\gamma }}
   \right)^{\alpha-1 }\log\left(1-e^{-\left(\beta  y_i\right)^{\gamma }}\right)\log\left(\beta
   y_i\right)
   \left(\beta  y_i\right)^{\gamma },\bigskip \\
  I_{\beta\beta} & =-\frac{n}{\beta ^2}-\gamma (\gamma -1)\sum _{i=1}^n  y_i^2 \left(\beta  y_i\right){}^{\gamma -2}+(\alpha -1)
  \gamma (\gamma -1)\sum _{i=1}^n y_i^2\frac{e^{-\left(\beta  y_i\right){}^{\gamma }}\left(\beta  y_i\right){}^{\gamma -2}}
  {1-e^{-\left(\beta  y_i\right){}^{\gamma }}}\medskip \\
   & ~~-(\alpha -1)\gamma ^2\sum _{i=1}^n  y_i^2\frac{e^{-2 \left(\beta  y_i\right){}^{\gamma }} \left(\beta  y_i\right){}^
   {2( \gamma -1)}}{\left(1-e^{-\left(\beta  y_i\right){}^{\gamma }}\right){}^2}-(\alpha -1)\gamma ^2 \sum _{i=1}^n  y_i^2
   \frac{e^{-\left(\beta  y_i\right){}^{\gamma }} \left(\beta  y_i\right){}^{2( \gamma-1) }}{1-e^{-\left(\beta  y_i\right){}^{\gamma }}}\medskip \\
   &~~+\alpha \theta \gamma (\gamma -1) \sum _{i=1}^n  y_i^2e^{-\left(\beta  y_i\right){}^{\gamma }} \left(1-e^{-
   \left(\beta  y_i\right){}^{\gamma }}\right){}^{\alpha -1} \left(\beta  y_i\right){}^{\gamma -2}\medskip \\
   & ~~-\alpha \theta  \gamma ^2
   \sum _{i=1}^n y_i^2e^{-\left(\beta  y_i\right){}^{\gamma }} \left(1-e^{-\left(\beta  y_i\right){}^{\gamma }}\right){}^{\alpha -1}
   \left(\beta  y_i\right){}^{2(\gamma -1)}\medskip \\
   & ~~+\alpha (\alpha -1)\theta \gamma ^2\sum _{i=1}^n y_i^2e^{-2 \left(\beta  y_i\right){}^{\gamma }}
   \left(1-e^{-\left(\beta  y_i\right){}^{\gamma }}\right){}^{\alpha -2} \left(\beta  y_i\right){}^{2(\gamma
   -1)},\bigskip \\
  I_{\beta\theta}&=\alpha  \gamma \sum _{i=1}^n  y_ie^{-\left(\beta
y_i\right){}^{\gamma }} \left(1-e^{-\left(\beta y_i\right){}^{\gamma
}}\right){}^{\alpha -1} \left(\beta y_i\right){}^{\gamma -1},\bigskip \\
  I_{\beta\gamma} & =-\sum _{i=1}^n y_i \left(\beta  y_i\right){}^{\gamma -1}-\gamma \sum _{i=1}^n
  y_i\log\left(\beta  y_i\right) \left(\beta  y_i\right){}^{\gamma -1}+(\alpha -1)
  \sum _{i=1}^n  y_i\frac{e^{-\left(\beta  y_i\right){}^{\gamma }}\left(\beta  y_i\right){}^{\gamma -1}}
  {1-e^{-\left(\beta  y_i\right)^{\gamma }}}\medskip \\
   & ~~+(\alpha -1)\gamma  \sum _{i=1}^n  y_i\frac{e^{-\left(\beta  y_i\right){}^{\gamma }}\log\left(\beta
   y_i\right)
   \left(\beta  y_i\right){}^{\gamma -1}}{1-e^{-\left(\beta  y_i\right){}^{\gamma }}}-(\alpha -1)\gamma
   \sum _{i=1}^n y_i\frac{e^{-2 \left(\beta  y_i\right){}^{\gamma }}\log\left(\beta  y_i\right) \left(\beta  y_i\right){}^{2
   \gamma -1}}{\left(1-e^{-\left(\beta  y_i\right){}^{\gamma }}\right){}^2}\medskip \\
   &~~-(\alpha -1)\gamma  \sum _{i=1}^n y_i\frac{e^{-\left(\beta  y_i\right){}^{\gamma }}\log\left(\beta
   y_i\right)
   \left(\beta  y_i\right){}^{2 \gamma -1}}{1-e^{-\left(\beta  y_i\right){}^{\gamma }}}+\alpha \theta  \sum _{i=1}^n y_ie^
   {-\left(\beta  y_i\right){}^{\gamma }} \left(1-e^{-\left(\beta  y_i\right){}^{\gamma }}\right){}^{\alpha -1} \left(\beta  y_i\right){}^{\gamma -1}\medskip \\
   &~~+\alpha \theta \gamma  \sum _{i=1}^n y_ie^{-\left(\beta  y_i\right){}^{\gamma }} \left(1-e^{-\left(\beta  y_i\right){}^{\gamma }}
   \right){}^{\alpha -1}\log\left(\beta  y_i\right) \left(\beta  y_i\right){}^{\gamma -1}\medskip \\
   &~~-\alpha \theta \gamma  \sum _{i=1}^n y_ie^
   {-\left(\beta  y_i\right){}^{\gamma }} \left(1-e^{-\left(\beta  y_i\right){}^{\gamma }}\right){}^{\alpha -1}\log
   \left(\beta  y_i\right)\left(\beta  y_i\right){}^{2 \gamma -1}\medskip \\
   &~~+\alpha (\alpha -1)\theta \gamma  \sum _{i=1}^n y_ie^{-2 \left(\beta  y_i\right){}^{\gamma }} \left(1-e^{-\left(\beta  y_i\right){}^
   {\gamma }}\right){}^{\alpha -2}\log\left(\beta  y_i\right)\left(\beta  y_i\right){}^{2 \gamma
   -1},
  \end{array}
\end{equation*}

\begin{equation*}
\begin{array}{ll}
I_{\theta\theta}&=\frac{ne^{-\theta }}{\left(e^{-\theta
}-1\right)^2}-\frac{n}{\theta ^2},\bigskip \\
 I_{\theta\gamma}&=\alpha \sum _{i=1}^n e^{-\left(\beta
y_i\right){}^{\gamma }} \left(1-e^{-\left(\beta y_i\right){}^{\gamma
}}\right){}^{\alpha -1}\log\left(\beta y_i\right) \left(\beta
y_i\right){}^{\gamma },\bigskip \\
  I_{\gamma\gamma} & =-\frac{n}{\gamma ^2}-\sum _{i=1}^n \log\left(\beta  y_i\right)^2 \left(\beta  y_i\right)^{\gamma }+(\alpha -1)
  \sum _{i=1}^n \frac{e^{-\left(\beta  y_i\right){}^{\gamma }}\log\left(\beta  y_i\right)^2 \left(\beta  y_i\right)^
  {\gamma }}{1-e^{-\left(\beta  y_i\right)^{\gamma }}}\medskip \\
   &~~-(\alpha -1) \sum _{i=1}^n \frac{e^{-2 \left(\beta  y_i\right){}^{\gamma }}\log\left(\beta  y_i\right)^2 \left(\beta  y_i\right)^
   {2 \gamma }}{\left(1-e^{-\left(\beta  y_i\right){}^{\gamma }}\right){}^2}-(\alpha -1) \sum _{i=1}^n \frac{e^{-\left(\beta  y_i\right)^{\gamma }}
    \log\left(\beta  y_i\right)^2 \left(\beta  y_i\right){}^{2 \gamma }}{1-e^{-\left(\beta  y_i\right)^{\gamma }}}\medskip  \\
   &~~+\alpha \theta  \sum _{i=1}^n e^{-\left(\beta  y_i\right)^{\gamma }} \left(1-e^{-\left(\beta  y_i\right)^{\gamma }}\right)^{\alpha-1 }
   \log\left(\beta  y_i\right)^2 \left(\beta  y_i\right){}^{\gamma }\medskip \\
   &~~-\alpha \theta  \sum _{i=1}^n e^{-\left(\beta  y_i\right)^{\gamma }}
   \left(1-e^{-\left(\beta  y_i\right)^{\gamma }}\right)^{\alpha-1 }\log\left(\beta  y_i\right)^2 \left(\beta  y_i\right)^
   {2 \gamma }\medskip  \\
   &~~+\alpha(\alpha -1) \theta\sum _{i=1}^n e^{-2 \left(\beta  y_i\right)^{\gamma }} \left(1-e^{-\left(\beta  y_i\right){}^{\gamma }}\right)^
   {\alpha-2 }\log\left(\beta  y_i\right)^2 \left(\beta  y_i\right)^{2 \gamma
   }.
\end{array}
\end{equation*}

\end{document}